\documentclass[journal,twoside,web]{ieeecolor}
\usepackage{generic}
\usepackage{cite}
\usepackage{amsmath,amssymb,amsfonts}
\usepackage{algorithmic}
\usepackage{gensymb}
\usepackage{graphicx}
\usepackage{textcomp}

\usepackage[hyphens]{url}
\usepackage[hidelinks]{hyperref}
\usepackage[hyphenbreaks]{breakurl}
\usepackage{mathtools}
\DeclarePairedDelimiter{\norm}{\lVert}{\rVert} 

\DeclareMathAlphabet{\mbf}{OT1}{ptm}{b}{n}

\def\BibTeX{{\rm B\kern-.05em{\sc i\kern-.025em b}\kern-.08em
    T\kern-.1667em\lower.7ex\hbox{E}\kern-.125emX}}
\begin{document}

%
%
%
%
%
%
%
\def \myJournal {IEEE Transactions on Biomedical Engineering}
\def \myDoi {10.1109/TBME.2022.3179655}
\def \myPaperSiteName {IEEE Xplore}
\def \myPaperSiteLink {https://ieeexplore.ieee.org/document/9786847}
\def \myYear {2022}
\def \myPaperCitation{A. Bianco, R. Zonis, A.-M. Lauzon, J. R. Forbes and G. Ijpma, ``System Identification and Two-Degree-of-Freedom Control of Nonlinear, Viscoelastic Tissues," in \textit{IEEE Transactions on Biomedical Engineering}, doi: 10.1109/TBME.2022.3179655.}


\begin{figure*}[t]

\thispagestyle{empty}
\begin{center}
\begin{minipage}{6in}
\centering
This paper has been accepted for publication in \emph{\myJournal}. 
\vspace{1em}

This is the author's version of an article that has, or will be, published in this journal or conference. Changes were, or will be, made to this version by the publisher prior to publication.
\vspace{2em}

\begin{tabular}{rl}
DOI: & \myDoi\\
\myPaperSiteName: & \texttt{\myPaperSiteLink}
\end{tabular}

\vspace{2em}
Please cite this paper as:

\myPaperCitation

\vspace{15cm}
\copyright \myYear \hspace{4pt}IEEE. Personal use of this material is permitted. Permission from IEEE must be obtained for all other uses, in any current or future media, including reprinting/republishing this material for advertising or promotional purposes, creating new collective works, for resale or redistribution to servers or lists, or reuse of any copyrighted component of this work in other works.

\end{minipage}
\end{center}
\end{figure*}
\newpage
\clearpage
\pagenumbering{arabic} 

\title{System Identification and Two-Degree-of-Freedom Control of Nonlinear, Viscoelastic Tissues}
\author{Amanda Bianco, Raphael Zonis, Anne-Marie Lauzon, James Richard Forbes, and Gijs Ijpma
\thanks{Manuscript received December 7, 2021; revised May 21, 2022; accepted May 22, 2022. This work was supported by the Natural Sciences and Engineering Research Council of Canada and Aurora Scientific Inc.~through the Collaborative Research and Development grant CRDPJ-519929-17 for the Lauzon Lab, and by the Fonds de recherche - Nature et technologies Master's Research Scholarship awarded to Amanda Bianco. \textit{(Corresponding author: A-M.~Lauzon)}}
\thanks{A.~Bianco was with the Department of Biomedical Engineering and Meakins-Christie Laboratories, McGill University, Montreal, QC, Canada. She is now with the Faculty of Medicine and Health Sciences, McGill University, Montreal, QC, Canada.}
\thanks{R.~Zonis was with the Department of Mechanical Engineering, McGill University, Montreal, QC, Canada. He is now with the Department of Mechanical Engineering, Massachusetts Institute of Technology, Cambridge, MA, USA.}
\thanks{A-M.~Lauzon is with the Department of Medicine, Department of Biomedical Engineering and Meakins-Christie Laboratories, McGill University, Montreal, QC, Canada (e-mail: anne.lauzon@mcgill.ca).}
\thanks{J.~R.~Forbes is with the Department of Mechanical Engineering, McGill University, Montreal, QC, Canada.}
\thanks{G.~Ijpma is with the Meakins-Christie Laboratories, McGill University, Montreal, QC, Canada.} 
\thanks{This article has supplementary downloadable material available at https://doi.org/10.1109/TBME.2022.3179655, provided by the authors.}
\thanks{Digital Object Identifier 10.1109/TBME.2022.3179655}}

\maketitle
\begin{abstract}
\textit{Objective:} This paper presents a force control scheme for brief isotonic holds in an isometrically contracted muscle tissue, with minimal overshoot and settling time to measure its shortening velocity, a key parameter of muscle function. \textit{Methods:} A two-degree-of-freedom control configuration, formed by a feedback controller and a feedforward controller, is explored. The feedback controller is a proportional-integral controller and the feedforward controller is designed using the inverse of a control-oriented model of muscle tissue. A generalized linear model and a nonlinear model of muscle tissue are explored using input-output data and system identification techniques. The force control scheme is tested on equine airway smooth muscle and its robustness confirmed with murine flexor digitorum brevis muscle. \textit{Results:} Performance and repeatability of the force control scheme as well as the number of inputs and level of supervision required from the user were assessed with a series of experiments. The force control scheme was able to fulfill the stated control objectives in most cases, including the requirements for settling time and overshoot. \textit{Conclusion:} The proposed control scheme is shown to enable automation of force control for characterizing muscle mechanics with minimal user input required. \textit{Significance:} This paper leverages an inversion-based feedforward controller based on a nonlinear physiological model in a system identification context that is superior to classic linear system identification. The control scheme can be used as a steppingstone for generalized control of nonlinear, viscoelastic materials.
\end{abstract}




\section{Introduction}
\IEEEPARstart{M}{uscle} mechanics measurements provide key insights into the contractile and non-contractile properties of healthy and diseased muscles. Additionally, these measurements can be used to assess the response to drugs, biological mediators and/or environmental factors. While crude mechanics measurements can be performed \textit{in vivo}, isolated muscle bundles need to be studied \textit{in vitro} in order to obtain accurate control over contractile stimulation in addition to direct measurement and control of length and force. \textit{In vitro} muscle mechanics measurements can be done at the cell or tissue level. At the cell level, techniques like traction force microscopy and collagen gel constructs can directly measure the forces that cultured muscle cells exert on a substrate \cite{Munevar, Tolic, Matsumoto}. However, studies show that individual muscle cells are prone to losing their contractile phenotype when studied in cell culture \cite{Chamley-Campbell, Arafat, Huber}. To better characterize \textit{in vivo} muscle mechanics, freshly dissected, intact muscle tissue can be used directly as a higher fidelity model.

To measure the force generated by a muscle tissue, one end of the tissue is attached to a force transducer, allowing direct force readout. However, a number of key parameters of muscle contractility require direct control of the force. One such parameter is the muscle's rate of shortening, otherwise known as the shortening velocity. The shortening velocity at no load, also known as the unloaded shortening velocity, is believed to be a direct correlate of the molecular level mechanics of the muscle's contractile elements \cite{Seow}. One way to measure the shortening velocity of muscle tissue \textit{in vitro} is via quick-release experiments in which the tissue is suspended between a force transducer and a length actuator and subjected to either a chemical or an electrical stimulus to induce contraction \cite{Seow}. During an isometric contraction of the muscle tissue, a sudden switch to isotonic force control is applied. The level of force to which the tissue is controlled will determine the rate at which the tissue shortens, i.e., the shortening velocity. As unloaded shortening velocity cannot be measured directly in an actively controlled tissue with no compressive strength, a series of different force level quick-release protocols, which are referred to as force clamps, are used to find the unloaded shortening velocity by extrapolating the hyperbolic curve that is fit to the data \cite{Seow2}. By the nature of the fit, the estimation of the unloaded shortening velocity is strongly dependent on force clamps closest to a purely unloaded state. In addition to acquiring shortening velocity measurements and an estimate of the unloaded shortening velocity, force clamps can also provide insight into other mechanical parameters such as elastic recoil. Force clamps were traditionally performed by attaching physical loads to tissue samples \cite{McMahon}, but are now performed with simulated loads via digital force control \cite{Matusovsky, Matusovsky2, Bullimore2, Luo, Ijpma}. 

The literature in the field of tissue mechanics does not provide much detail regarding the force control scheme used for force clamps since researchers often use commercial equipment with control schemes that are unpublished and require manual tuning \cite{Matusovsky, Matusovsky2, Bullimore2, Luo, Ijpma}. The authors of \cite{Sabourin} present the design of a feedback system involving proportional-integral (PI) control for isotonic studies of smooth muscle \textit{in vitro}, but manual tuning is required. As a consequence of intra- and inter-tissue variability, coupled with the dynamic, nonlinear, viscoelastic nature of muscle tissue, it is difficult to design a static, linear feedback controller that can successfully perform force clamps in all test cases. This lack of versatility makes the performance of force clamps highly cumbersome, and precludes the throughput necessary to perform high volume studies, as well as the possibility of complete automation.

The contributions of this paper include leveraging a nonlinear physiological model in a system identification (ID) context that is superior to classic linear system ID and the implementation of an inversion-based feedforward controller to help automate force clamps for measuring muscle mechanics. While specifically applied to the determination of shortening velocity here, the approach can be generalized to any force control of nonlinear, viscoelastic materials. This paper is structured as follows: Section \ref{sec: materials_and_methods} describes the experimental setup and protocols needed to test the force control scheme, Section \ref{sec: force_control_scheme} details the force control scheme along with the control objectives and the system ID process, Section \ref{sec: results} presents the experimental results and Section \ref{sec: discussion} discusses the experimental results with respect to the control objectives. 


\section{Materials and Methods}
\label{sec: materials_and_methods}

\subsection{Muscle Tissue Procurement and Preparation}

The force control scheme was developed and tested in experiments with equine airway smooth muscle (ASM) and its robustness was confirmed with murine flexor digitorum brevis (FDB) muscle, which is a skeletal muscle. Equine trachea segments were obtained from a slaughterhouse, after which the ASM with epithelium and connective tissue was dissected out and cryostored according to \cite{Ijpma2}. On the day of the experiment, ASM tissues were dissected free from the epithelium and connective tissue in Ca$^{2+}$-free Krebs-Henseleit solution (refer to Section 1.1 of the Supplemental Material for solution compositions). Aluminum foil clips were then attached to either end of each tissue, which was approximately 3$\times$1$\times$0.5 mm. The ASM tisues that were not immediately used were pinned to pieces of silicone at in situ length and kept in Dulbecco's modified Eagle's medium with 2$\%$ heat-inactivated fetal bovine serum at 37$^{\circ}$C in an incubator for no more than 3 days. Murine tissues were procured from waste tissue from mice used for unrelated research. All procedures with murine FDB muscle were approved by the McGill University Animal Care Committee and performed in accordance with the guidelines of the Canadian Council on Animal Care (protocol 8082 approved in June 2019). The FDB muscles were dissected from the hind paws according to \cite{Demonbreun, Keire}, removing as much of the surrounding tissue as possible. Aluminum foil clips were then attached to either end of each FDB muscle, which was approximately 5$\times$1$\times$1~mm. FDB muscles that were not immediately used were preserved in the same manner as ASM.

\subsection{Tissue Bath Setup}
\label{sec: tissue_bath_setup}

\subsubsection{Hardware}

\begin{figure}[]
  \centering
  \includegraphics[width=\columnwidth]{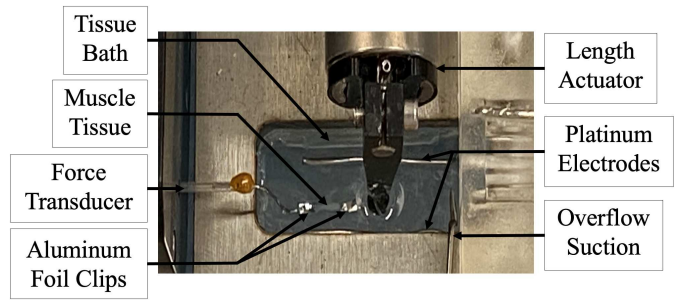}
    \setlength{\abovecaptionskip}{-5pt}
  \caption{Tissue bath setup, where the muscle tissue is ASM.}
  \label{fig: bianc1}
\end{figure}

The tissue bath setup includes a force transducer (model 400A from Aurora Scientific Inc.) and length actuator (model 322C from Aurora Scientific Inc.) on either side of a tissue bath (Fig.~\ref{fig: bianc1}). The resonant frequency of the force transducer and length actuator are approximately 140~Hz and 2~kHz, respectively, both of which are greater than the assumed bandwidth of muscle tissue, e.g., 100 Hz or less based on the frequency spectrum of the inputs explored in \cite{Kirsch}. The voltage-to-force conversion factor of the force transducer was 5~mN/V and the voltage-to-length conversion factor of the length actuator was 0.3~mm/V. Fluid flow was controlled by two peristaltic pumps, one for relaxing solution and one for contracting solution, at 1~mL/min into a total bath volume of 2.4~mL, with overflow suction to a vacuum. Platinum electrodes, which were connected to an electrical stimulator (model 701C from Aurora Scientific Inc.), were placed in the tissue bath on either side of the muscle tissue for electric field stimulation (EFS). 

Data acquisition, force control, peristaltic pump control and EFS control were achieved via a FPGA-controlled USB-7845 system from National Instruments.

\subsubsection{Software}

LabVIEW code, uploaded to the FPGA system, controlled the tissue bath's hardware. MATLAB was primarily used for data analysis, but MATLAB scripts were executed within the LabVIEW environment of the host computer primarily for model estimation/validation and feedforward generation, all of which are described in Section \ref{sec: force_control_scheme}.

\subsection{Experimental Protocols}

All experiments were conducted at room temperature $\pm$1$\degree$C. The tissue bath was filled with Ca$^{2+}$-free Krebs-Henseleit solution and a muscle tissue was mounted between the force transducer and length controller (Fig.~\ref{fig: bianc1}). Once the tissue was mounted, it was equilibrated with repeated contractions induced either by 10$^{-5}$~M methacholine solution for ASM or EFS for FDB muscle (refer to Section 1 of the Supplemental Material for more details on chemical and electrical stimulation parameters). After model estimation/validation and feedforward generation, force clamps at 6 levels, e.g., 5$\%$, 7$\%$, 10$\%$, 20$\%$, 40$\%$ and 80$\%$ of the steady state contractile force achieved directly prior to force clamp initiation, were applied in random order at 1-min intervals, all within one contraction for ASM and in separate contractions for FDB muscle on account of the short maximal duration of EFS. Note that 5$\%$ and 80$\%$ are the largest and smallest amplitude force clamp levels, respectively, since the reference contractile force was dropped by 95$\%$ for the former and by 20$\%$ for the latter. To test the repeatability of the force control scheme, force clamps were repeated within and across contractions of the same contractile force for ASM. As for FDB muscle, force clamps were repeated across contractions of the same as well as different contractile force. The repeatability and robustness of the force control scheme depends on that of the system ID process described in Section \ref{sec: system_ID}, specifically model estimation. Thus, it was important to assess whether the same model could be estimated each time this process was repeated and if the same estimate could then be used for each force clamp level within or across contractions of the same or different contractile force. This assessment was mainly done for the selected nonlinear model. Refer to Section 1 of the Supplemental Material for the details regarding the experimental protocols.

\section{Force Control Scheme}
\label{sec: force_control_scheme}

\subsection{Control Objectives}

Different experiments have different requirements for settling time and overshoot based on the research needs. The maximum settling time is set here to 60~ms so that shortening velocity measurements can take place as early as 60~ms into the force clamp. The allowable overshoot is less than 3$\%$ of the contractile force just prior to the force clamp, specifically for the larger amplitude force clamp levels, e.g., those of 10$\%$ or less, to avoid the highly nonlinear tissue mechanics near slack length. The force control scheme also needs to be repeatable, robust and require little to no inputs and supervision from the user, which are especially important objectives for long-term automated tissue studies. It is important to note that shortening velocity is neither measured nor evaluated in this paper because there do not exist true shortening velocity values against which measurements can be compared. This is why the control objectives are focused on tracking the reference signal via the control parameters of settling time and overshoot, with the emphasis being on the larger amplitude force clamp levels for accurate extrapolation of the unloaded shortening velocity.

\subsection{Control Configuration}

\subsubsection{Overview}
\begin{figure}[]
  \centering
  \includegraphics[width=\columnwidth]{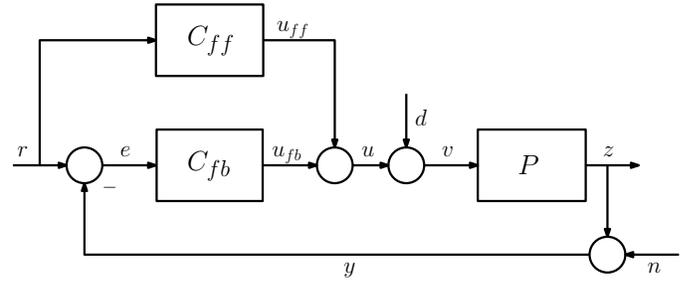}
    \setlength{\abovecaptionskip}{-5pt}
  \caption{2DOF control configuration, where $d$ is the disturbance, $v$ is the plant's input, $z$ is the plant's output and $n$ is the measurement noise.}
  \label{fig: bianc2}
\end{figure}

The implemented two-degree-of-freedom (2DOF) control configuration \cite{Qiu} is shown in Fig.~\ref{fig: bianc2}. It comprises the plant $P$, feedback controller $C_{fb}$ and feedforward controller $C_{ff}$. The difference between the reference signal $r$ and measured output $y$ is the error $e$
\begin{equation}
e = r - y,
\label{tracking error}
\end{equation}
which serves as the input to the feedback controller. The reference signal also serves as the input to the feedforward controller. The total control effort $u$ is given by
\begin{equation}
u = u_{ff} + u_{fb},
\label{2DOF control effort}
\end{equation}
where $u_{ff}$ is the control effort of the feedforward controller and will be referred to as the feedforward signal and $u_{fb}$ is the control effort of the feedback controller. The feedforward and feedback controllers can be designed together or independently. In this work, for simplicity as well as to leverage system ID techniques and information provided by physiological models, they are designed independently.

\subsubsection{Reference Signal}

The reference signal consists of a 600~ms step to the desired force clamp level scaled by the reference contractile force. This reference force is determined by averaging 100 samples of force reading prior to initiating the force clamp. The reference signal is ramped down over 20~ms to help manage the hardware delay between the input and output of the plant as well as the resonance of the force transducer. A sample reference signal is shown in Fig.~\ref{fig: bianc3} and the normalized reference signal for each force clamp level is shown in Fig.~S1 of the Supplemental Material.

\begin{figure}[]
  \centering
  \includegraphics[width =\columnwidth]{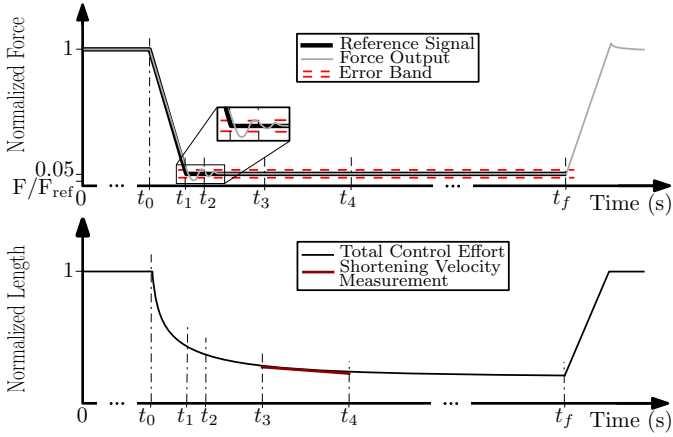}
    \setlength{\abovecaptionskip}{-5pt}
  \caption{Example of normalized force and length traces of a 5$\%$ force clamp with overshoot, where F$_{\text{ref}}$ is the reference contractile force, 0.05 is the final normalized force reached by the reference signal and F is the force measured by the force transducer. The start of the force clamp is $t_{0} = 0.08$~s, the ramp down of the reference signal ends at $t_{1} = 0.1$~s, the force output settles between the error band as of $t_{2}$ and the shortening velocity is estimated from $t_{3} = 0.14$~s to $t_{4} = 0.19$~s. At $t_{f} = 0.6$~s, the force control is turned off and the length is ramped to the reference length.}
  \label{fig: bianc3}
\end{figure}

\subsubsection{Plant}

The plant comprises a series combination of the length actuator, muscle tissue and force transducer. The plant is a single-input, single-output system, where the input is length in volts and the output is force in volts. The muscle tissue is assumed to be the main contributor to the dominant dynamics of the plant, particularly during the phase of the force clamp in which the tissue's shortening velocity is measured since, in this phase, the effects of resonance of the force transducer and length controller in response to the initial rapid length change and the ensuing interaction with the feedback controller have subsided. This is why, in Section \ref{sec: system_ID}, modelling efforts of the plant are focused on the muscle tissue itself. 

\subsubsection{Feedback Controller}

Feedback control helps maintain good tracking performance in the presence of disturbances and model uncertainty \cite{Qiu, Astrom}. The feedback controller is a PI controller, for which the control effort $u_{fb}(t)$ is the sum of two terms
\begin{equation}
u_{fb}(t) = k_{p}e(t) + k_{i}\int_{0}^{t} e(\tau) d\tau,
\end{equation}
\noindent
where $k_{p}$ is the proportional gain, $k_{i}$ is the integral gain and $e(t)$ is given by \eqref{tracking error} \cite{Astrom}.  PI controllers are ideal for tracking step reference signals because they lead to zero steady-state error thanks to the integral action \cite{Astrom}. The step nature of the force clamp reference signal is therefore the main reason for choosing the PI controller as the feedback controller. Prior experiments focused on finding acceptable gains showed good settling time and overshoot results with $k_{p}$~$=$~$0$ and $k_{i}$~$=$~$0.04$~s$^{-1}$.

\subsubsection{Feedforward Controller}
In theory, if a model perfectly describes the input-output relationship of a plant and the initial conditions are perfectly known, an inversion-based feedforward controller, which at its simplest is the inverse model of the plant, perfectly tracks the reference signal without the need for a feedback controller. However, no such model exists in practice and a feedback controller is also necessary. Knowing that the PI controller is linear and the plant is nonlinear suggests that a nonlinear feedforward controller could realize the best performance, but this will depend on the degree of nonlinearity of the plant. Using nonlinear feedforward with linear feedback is a standard approach in the field of robotics \cite{Astrom}. The feedforward controller is generated using a model of the plant dynamics that is estimated in real time via experimental input-output data and system ID techniques (Section \ref{sec: feedforward_controller}). 

\subsection{System ID of Input-Output Data}
\label{sec: system_ID}

\subsubsection{Models}
\label{sec: Models}

As the rapid dynamics of muscle tissue pose the greatest control challenge, modelling efforts for the feedforward controller focused predominantly on reliably capturing these dynamics. It is assumed that the contractile dynamics of the tissue change slowly relative to the time scale of a typical force clamp, allowing the tissue to be modelled as a pseudo-static, viscoelastic material, where actively changing tissue properties can be assumed to be constant over the course of a typical force clamp. The aim is to find a model that optimizes the ability to capture the muscle tissue's properties, while keeping the number of free parameters low. This maximizes feedforward effort thus minimizing feedback effort, while also ensuring robustness and repeatability. The choice of model for the feedforward design can be reduced to the choice between linear and nonlinear models. A generalized linear model and a minimalistic nonlinear model were used to assess whether the nonlinearity is essential to achieving the control objectives.

\begin{figure}[]
  \centering
  \includegraphics[width = \columnwidth,  trim={1cm 0 4cm 1cm}, clip]{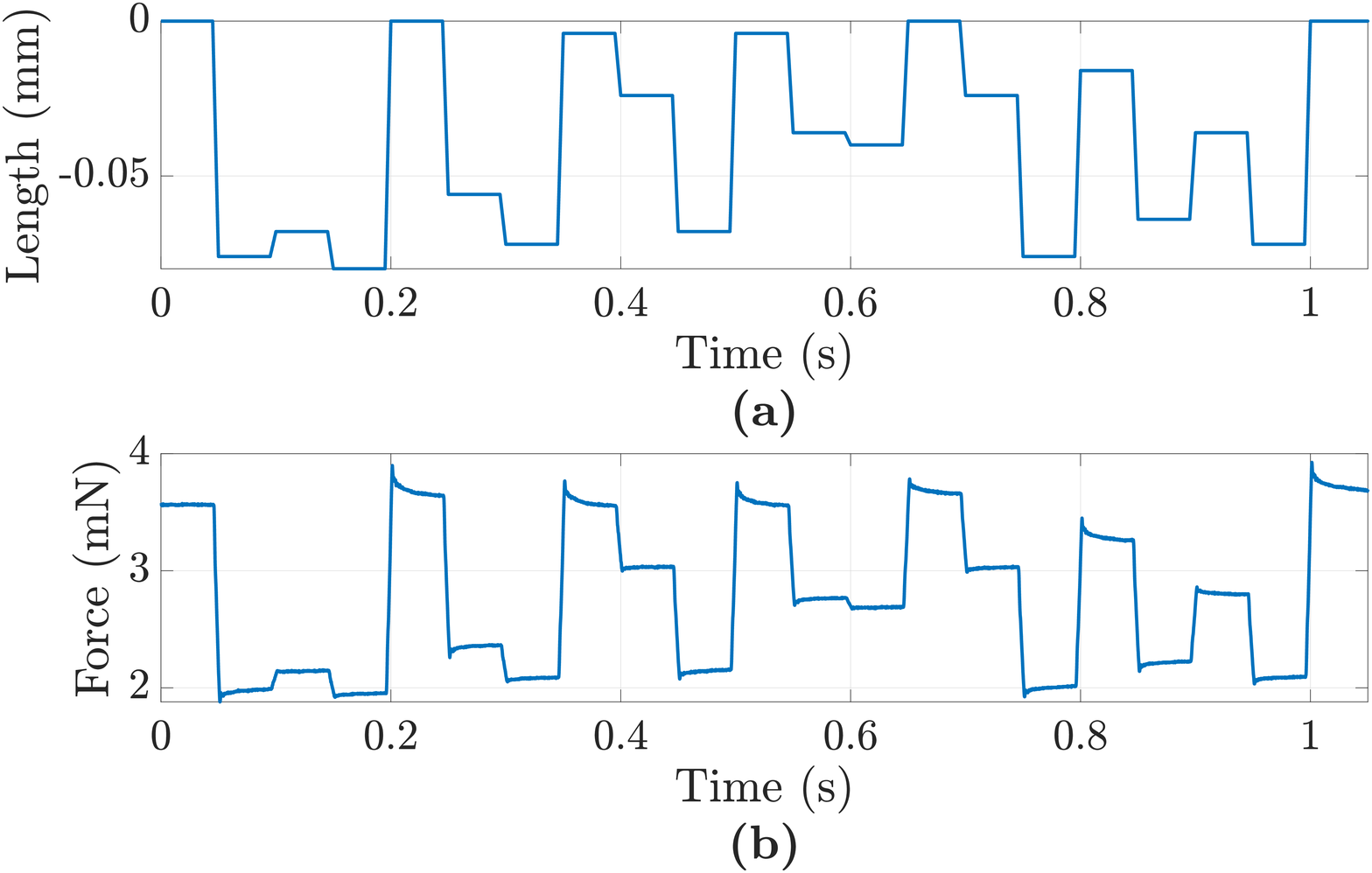}
    \setlength{\abovecaptionskip}{-10pt}
  \caption{Sample (a) input and (b) corresponding output to estimate the nonlinear model.}
  \label{fig: bianc4}
\end{figure}

The first model, which will be referred to as the linear model, is a generalized linear model in the form of a biproper transfer function of order $n$ given by
\begin{equation}
H(z) = \dfrac{\alpha_{n} + \alpha_{n-1}z^{-1} + \ldots + \alpha_{1}z^{-n+1} + \alpha_{0}z^{-n}}{1 + \beta_{n-1}z^{-1} + \ldots + \beta_{1}z^{-n+1} + \beta_{0}z^{-n}}, \label{eq: model_1}
\end{equation}

\noindent
where $\alpha_{i}$ and $\beta_{i}$ are the parameters to be estimated and the initial conditions are assumed to be quiescent (refer to Section 2.2.1 of the Supplemental Material for the derivation) \cite{Hespanha}. This model structure was informed by two linear, viscoelastic models that were previously explored: the generalized Maxwell model \cite{Gutierrez} and the Golla-Hughes-McTavish model \cite{McTavish}, both of which yield biproper transfer functions. Systems that can be represented by biproper transfer functions are known to exhibit feedthrough, i.e., a finite, non-zero gain can be observed at infinitely high input frequencies. In the case of muscle tissue, feedthrough is observed as a step change in length results in an immediate, finite change in force (see Fig.~\ref{fig: bianc4}).

\begin{figure}[]
\centering
   \includegraphics[width=\columnwidth]{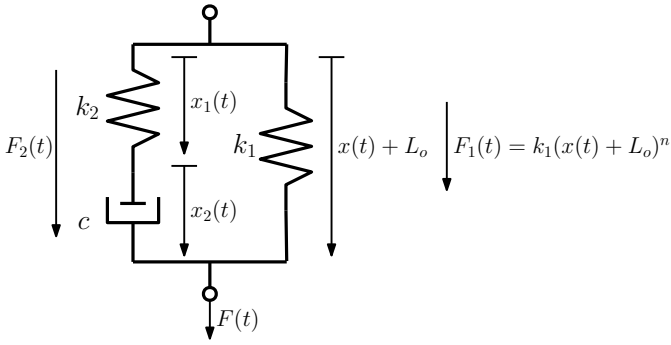}
     \setlength{\abovecaptionskip}{-5pt}
   \caption{First-order Maxwell model with a zeroth-order, nonlinear spring with power law dynamics.}
 \label{fig: bianc5}
\end{figure}

The second model, which will be referred to as the nonlinear model, is a first-order Maxwell model \cite{Fung} with a zeroth-order, nonlinear spring (Fig.~\ref{fig: bianc5}). The zeroth-order spring is chosen to be nonlinear because it is responsible for the steady-state force value following a step change in length. More specifically, it takes the form of a power law because a power law relationship was observed when plotting elastic recoil length versus force clamp values for a given muscle tissue using data from \cite{Ijpma}. The ordinary differential equation of this model is given by
\begin{equation}
\begin{split}
\dot{F}(t) = (k_{2} + k_{1}n(x(t)+ L_{0})^{n - 1})\dot{x}(t) - \dfrac{k_{2}}{c}F(t)\\
 + \dfrac{k_{1}k_{2}}{c}(x(t) + L_{o})^{n}, \label{eq: model_2}
\end{split}
\end{equation}
where spring constants $k_{1}$ and $k_{2}$, viscous damping coefficient $c$ and exponent $n$ are to be estimated (refer to Section 2.2.2 of the Supplemental Material for the derivation). The muscle tissue's projected unstrained length $L_{o}$ is given by
\begin{equation}
L_{o} = \left(\dfrac{F(0)}{k_{1}}\right)^{\!1/n}.
\end{equation}
\noindent

\subsubsection{Designing Input-Output Data for Model Estimation and Validation}

Separate datasets are used for model estimation and validation as dynamic tissue remodelling can cause past length changes to affect future force responses. The input-output data need to capture the dynamics relevant to the control of force clamps, which combine a rapid step-like phase with a maintained, slower phase. To capture these dynamics, the input signal is designed using ramped step length changes of a range of amplitudes. The upper bound of the input is set to 0 V, which corresponds to the length at which the tissue is isometrically contracted, i.e., the reference length, such that the muscle tissue is never stretched beyond this length to avoid tissue damage. The lower bound of the input is set to the upper bound minus 2$\%$ of the muscle tissue's reference length to avoid slack muscle tissue which is problematic for system ID. Sample input-output data to estimate the nonlinear model are shown in Fig.~\ref{fig: bianc4}. Refer to Figs.~S2 and S3 of the Supplemental Material for sample input-output data to estimate the linear model and sample input-output data to validate both models, respectively.

The input-output data are preprocessed prior to model estimation and validation in order to yield the best results for a given model. There is a delay of approximately 1 ms between the input and output data caused by response delays of the length actuator and force transducer, which cannot be captured by the models. The delay is removed by shifting the output data relative to the input data for both model estimation and validation. For the linear model, since \eqref{eq: model_1} is formed on the assumption that the initial conditions are quiescent, the offset is removed from the output data for model estimation and validation so that the starting force is 0~V. Also, in order for the computation of the least squares estimate (Section \ref{model_estimation}) to be numerically well-conditioned, the input-output data for model estimation are normalized to the range $\left[-1, 1\right]$. Prior to model validation, the scaling is reversed to obtain the original transfer function $H(z)$. For the nonlinear model, the output data for model estimation and validation are filtered using a moving average filter with a window size of 20 samples in order to smooth out any resonance and measurement noise from the force transducer. 

\subsubsection{Model Estimation}
\label{model_estimation}

The parameters of the linear model are estimated for order $n \leq 3$ using MATLAB's \texttt{tfest} \cite{MathWorks}, which involves least squares in its default settings. Although higher order models have the potential to improve the quality of fit, they give rise to more variability in the parameter estimates as a result of overfitting. The resulting discrete-time transfer function is then converted to continuous time by \texttt{tfest}. 

The parameters of the nonlinear model are estimated via constrained optimization using MATLAB's \texttt{fmincon} \cite{MathWorks2} in its default settings. Equation \eqref{eq: model_2} is solved at every iteration of \texttt{fmincon} to minimize the cost function given by the normalized root mean square error (NRMSE), 
\begin{equation}
\textrm{NRMSE} = \sqrt{\dfrac{\dfrac{1}{N}\sum_{i = 0}^{N}(y_{i} - \hat{y}_{i})^{2}}{\dfrac{1}{N}\sum_{i = 0}^{N}y_{i}^{2}}},
\label{NRMSE}
\end{equation}
\noindent
where $\boldsymbol{y}$ is the observed output, $\hat{\boldsymbol{y}}$ is the simulated output and the error or the residual $\boldsymbol{r}$ is defined as $\boldsymbol{r} = \boldsymbol{y} - \hat{\boldsymbol{y}}$. The NRMSE is equal to 0 when there is complete overlap between the simulated and observed outputs. In this case, the observed output is the preprocessed observed output. Equation \eqref{eq: model_2} is solved using the classical fourth-order Runge-Kutta method \cite{Heath} since a fixed time step is needed for \texttt{fmincon} to converge. The initial guess and constraints for each of the free parameters of the nonlinear model are provided in Section 2.2.2 of the Supplemental Material.

\subsubsection{Model Validation}

Regardless of the type of models used for this application, NRMSE is used to assess 1) the quality of fit between the simulated output and the preprocessed observed output for model validation and 2) the quality of fit between the total control effort and the feedforward signal for each force clamp level. For the linear model, NRMSE is also used to select the model order that results in the best fit, i.e., the model order for which the NRMSE is closest to 0 using the simulated output and the preprocessed observed output for model validation.

\subsubsection{Feedforward Controller}
\label{sec: feedforward_controller}

Once the continuous-time transfer function of the linear model is obtained, it is inverted by switching the numerator and denominator in order to generate the transfer function of the feedforward controller. However, if prior to the inversion the numerator contains nonminimum phase zeros, which are zeros in the open right half plane, they are first mirrored about the imaginary axis into the open left half plane \cite{Qiu}. This is done to ensure that the transfer function is bounded-input, bounded-output stable once inverted, i.e., all of its poles are in the open left half plane. Mirroring non-minimum phase zeros does not change the gain of the transfer function, but it changes its phase. The feedforward signal is then simulated using MATLAB's \texttt{lsim} \cite{MathWorks3} with the reference signal as the input to the feedforward controller. A median filter is applied to remove an initial spike in the response due to the drop in force since the linear model does not accurately estimate the feedthrough.

Once the parameters of the nonlinear model are estimated, they are substituted into the ordinary differential equation of the feedforward controller given by
\begin{equation}
\begin{split}
\dot{x}(t) = \dfrac{1}{(k_{2} + k_{1}n(x(t) + L_{o})^{n - 1})}(\dot{F}(t) + \dfrac{k_{2}}{c}F(t)\\
 - \dfrac{k_{1}k_{2}}{c}(x(t) + L_{o})^{n}), \label{eq: inverse_model_2}
\end{split}
\end{equation}
\noindent
which is the inverse input-output relationship of \eqref{eq: model_2} (refer to Section 2.2.2 of the Supplemental Material for the derivation). The feedforward signal is then simulated using the reference signal as the input to the feedforward controller.

\subsection{Data Analysis}

A sampling frequency of 10 kHz is used, which is more than the minimum required of ten times the assumed bandwidth \cite{Ljung} of muscle tissue \cite{Kirsch}. Prior to calculating the settling time, the force response is normalized by the reference contractile force. It is also filtered using a moving average filter with a window size of 20 samples; otherwise, the measurement noise affects the settling time calculation. As shown in Fig.~\ref{fig: bianc3}, the settling time is calculated as the time it takes for the force response to settle within an error band from the start of the force clamp at $t_{0}~=$~ 0.08~s. An error band of $\pm$0.25$\%$ of the starting force is chosen because it corresponds to less than a $\pm$20$\%$ deviation of the measured shortening velocity for a 5$\%$ force clamp, which is acceptable for the research needs in this study. A constant absolute magnitude error band is used as control effort is proportional to the absolute error and for higher force clamps, measurement noise would exceed a constant relative magnitude error band. Using Fig.~\ref{fig: bianc3} as an example, the overshoot $OS$ is calculated as 
\begin{equation}
OS = \dfrac{(1 - \frac{F}{F_{\text{ref}}}) - (1 - \frac{F_{\text{level}}}{100})}{1 - \frac{F_{\text{level}}}{100}} \times 100\%, \label{eq: percent_overshoot}
\end{equation}
where $F_{\text{level}}$ is the force clamp level. Note that an overshoot of 5$\%$ for a 5$\%$ force clamp corresponds to slack muscle tissue.

\section{Results}
\label{sec: results}
 
\subsection{System ID Results}

Comparison of the linear versus the nonlinear models, using the same input-output data, clearly shows better predictive power of the nonlinear model in open loop for each data point (Fig.~\ref{fig: bianc6}~(a)); sample output data for model validation is provided in Fig.~S4 of the Supplemental Material. This is further confirmed when comparing the quality of the feedforward signal generated from both models relative to the total control effort from the 2DOF control configuration, which uses the feedforward signal generated by the nonlinear model. Fig.~\ref{fig: bianc6}~(b) shows the NRMSE of this comparison for the 5$\%$ force clamps, while Fig.~S5 of the Supplemental Material shows a sample of the comparison for 5$\%$ and 80$\%$ force clamps. The larger the amplitude of the force clamp level, the greater the display of the muscle tissue's nonlinear, viscoelastic properties, which the linear model falls short in capturing as compared to the nonlinear model. These data clearly show that the nonlinear model performs better than the linear model in reducing the control effort of the PI controller. To confirm the validity of the force clamps that correspond to the total control efforts used to compute NRMSE, their settling time and overshoot values are provided in Table S1 of the Supplemental Material. From this point forward, the remaining results will only be presented with respect to the nonlinear model.

\begin{figure}[]
  \centering
  \includegraphics[width = \columnwidth,  trim={2cm 0 4cm 0}, clip]{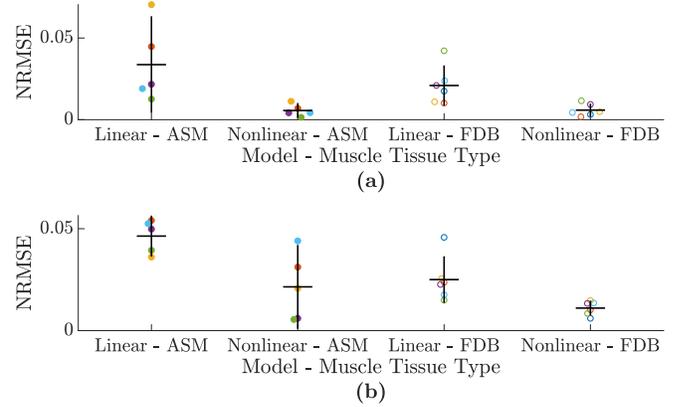}
    \setlength{\abovecaptionskip}{-10pt}
\caption{(a) NRMSE computed with the simulated output and the preprocessed observed output for model validation of both models for all muscle tissues. (b) NRMSE computed with the total control effort for a 5$\%$ force clamp and corresponding feedforward signal generated with either model for each muscle tissue. ASM tissues are represented by $\bullet$ and FDB muscle tissues are represented by $\circ$. The vertical black lines are 95$\%$ confidence intervals and the horizontal black lines are the mean values. For plotting purposes, each muscle tissue was assigned a colour that is used in all the remaining figures.}
  \label{fig: bianc6}
\end{figure}

\begin{figure}[]
  \centering
  \includegraphics[width = \columnwidth,  trim={0.8cm 0 4cm 0}, clip]{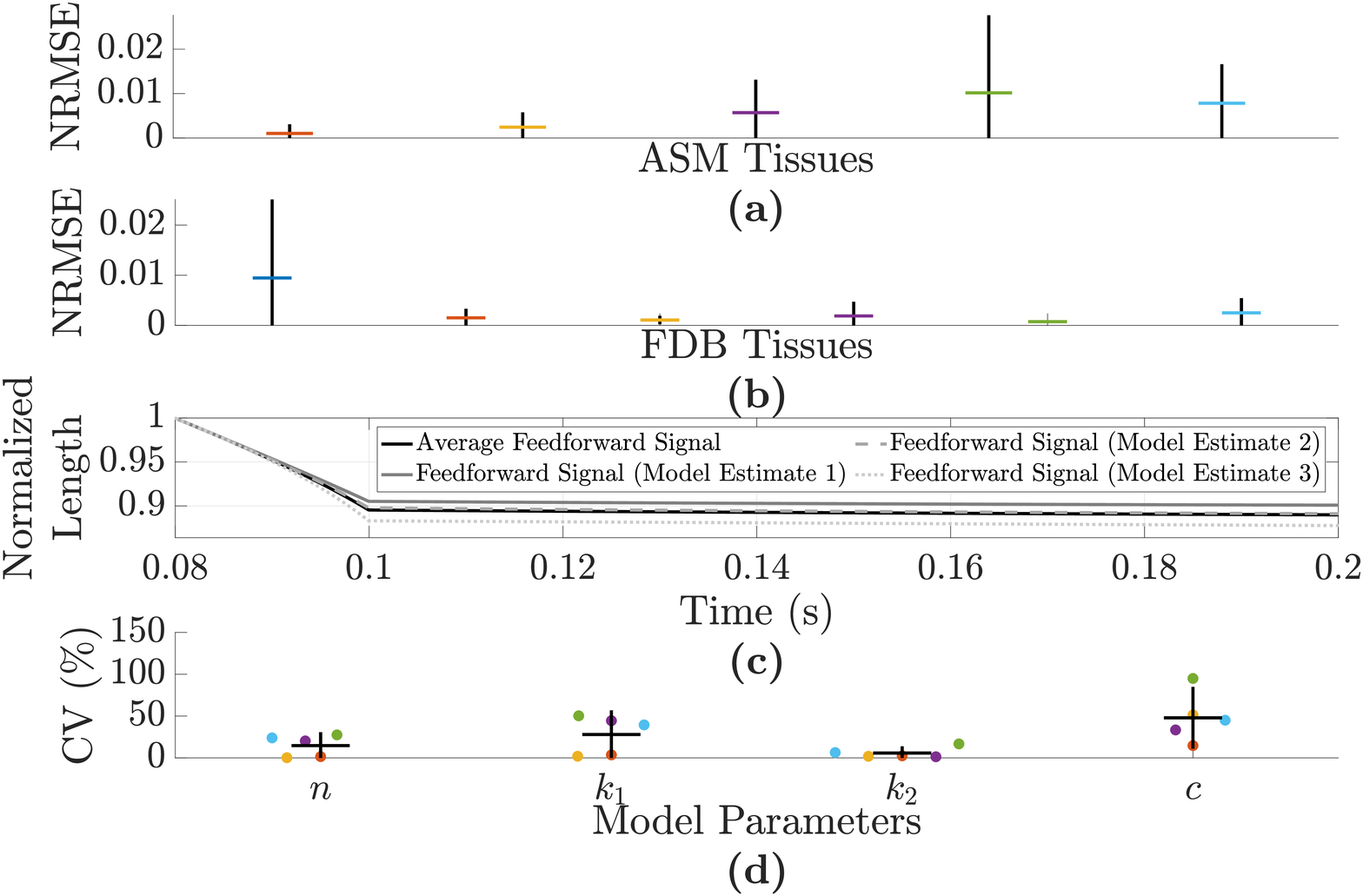}
    \setlength{\abovecaptionskip}{-5pt}
  \caption{(a) NRMSE computed with the average feedforward signal generated with the nonlinear model that is estimated in three consecutive rounds within the same contraction for each ASM tissue and each feedforward signal. (b) NRMSE computed with the average feedforward signal generated with the nonlinear model that is estimated in three consecutive contractions with the same contractile force for each FDB muscle and each feedforward signal.  (c) Average feedforward signal from $t =$~0.08~s to $t =$~0.2~s and corresponding feedforward signals for the ASM tissue with the greatest variance in (a). For (a)-(c), the feedforward signals correspond to a 5$\%$ force clamp. (d) Coefficient of variation, CV, for the parameter estimates of each model estimate in (a). The vertical black lines are 95$\%$ confidence intervals and the horizontal lines are the mean values.}
  \label{fig: bianc7}
\end{figure}

The repeatability of the estimation process for the nonlinear model was assessed by estimating this model three consecutive times within the same contraction for ASM (Fig.~\ref{fig: bianc7}~(a)) and across contractions with the same reference contractile force for FDB muscle (Fig.~\ref{fig: bianc7}~(b)). If the estimation process yields repeatable results for a tissue with a given reference contractile force, then the feedforward signal for this tissue for a given force clamp should be fairly consistent. However, it can be observed from the average NRMSE values presented in these figures that there is some variability. The corresponding feedforward signals of the ASM tissue with the greatest variance based on the 95$\%$ confidence intervals in Fig.~\ref{fig: bianc7}~(a) are displayed in Fig.~\ref{fig: bianc7}~(c). The variability can be observed in the parameter estimates within and across contractions for all muscle tissues and it manifests itself in the feedforward signals that are generated. The coefficient of variation for the parameters of each model estimate in Fig.~\ref{fig: bianc7}~(a) is shown in Fig.~\ref{fig: bianc7}~(d). Note that the coefficient of variation can get distorted for small sample means as is the case for $c$, $k_{2}$ and sometimes $k_{1}$. For either ASM or FDB muscle, the greatest variability is observed for $k_{1}$ and $c$. 

\begin{figure}[]
  \centering
  \includegraphics[width = \columnwidth,  trim={1.5cm 14cm 4cm 2cm}, clip]{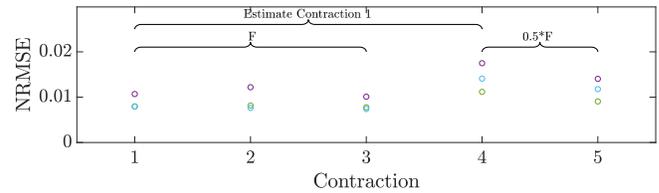}
  \setlength{\abovecaptionskip}{-15pt}
  \caption{NRMSE computed between the reference signal and force output for 5$\%$ force clamps with three FDB muscle tissues. Stimulation level was kept constant for the first three contractions and reduced to generate half the contractile force for the last two contractions. The same estimate for the nonlinear model was used for the first four contractions, with a new estimate for the fifth contraction at the lower contractile force. Control gains are kept constant.}
  \label{fig: bianc8}
\end{figure}

The quality of the feedforward signal generated by the nonlinear model for repeated force clamps at the same contractile force and at reduced contractile force in FDB muscle is shown in Fig.~\ref{fig: bianc8}. This was done in FDB muscle only because it is easier to reliably reduce the contractile force with EFS contractions. The NRMSE values in Fig.~\ref{fig: bianc8} are computed using the reference signal and force output for each 5$\%$ force clamp. Lower NRMSE values are achieved when the feedforward signal is generated using the model that is estimated at the same contractile force as the force clamp, which suggests that the system ID process may need to be repeated for large differences in contractile force. Table S2 of the Supplemental Material, which includes the corresponding values of settling time and overshoot, supports this observation and also suggests that a change in control gains may be required in addition to model re-estimation.

\subsection{Force Control Results}
    
\begin{figure}[]
  \centering
  \includegraphics[width = \columnwidth,  trim={1.5cm 0 4cm 0}, clip]{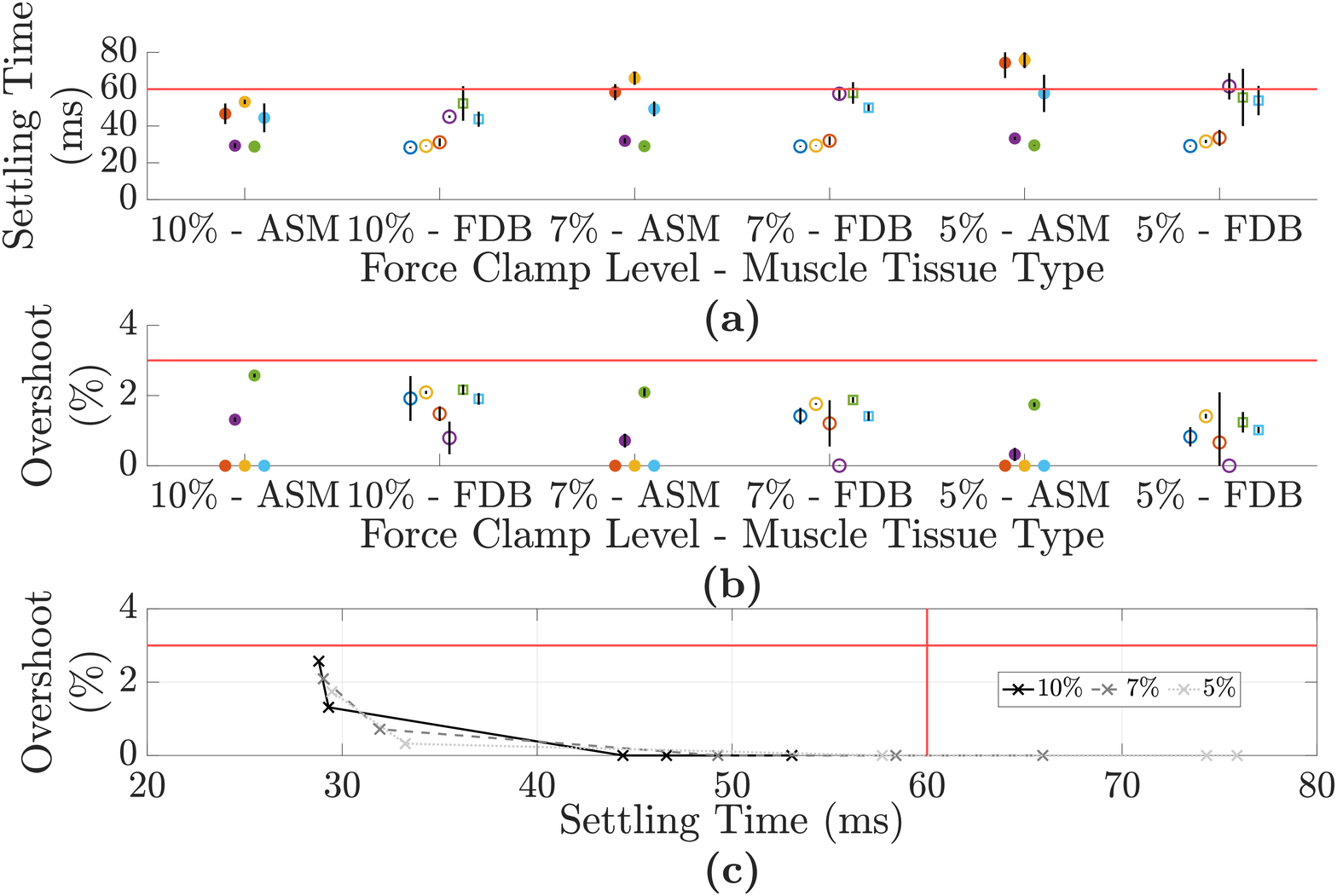}
    \setlength{\abovecaptionskip}{-5pt}
  \caption{(a) Settling time for repeated 10$\%$, 7$\%$ and 5$\%$ force clamps for ASM ($\bullet$) and FBD ($\circ$) muscle. (b) Overshoot for repeated 10$\%$, 7$\%$ and 5$\%$ force clamps for ASM and FBD muscle. Note that square symbols represent cases in (a) and (b) where $k_{i}$~$= 0.02$~s$^{-1}$. (c) Average overshoot versus average settling time using results from (a) and (b), where each data point for each force clamp level is a different ASM tissue. The red lines in (a)-(c) correspond to the requirements for settling time and overshoot. The vertical black lines are 95$\%$ confidence intervals.}
  \label{fig: bianc9}
\end{figure}

The average settling time and average overshoot are shown for repeated 10$\%$, 7$\%$ and 5$\%$ force clamps for ASM and FDB muscle in Fig.~\ref{fig: bianc9}. A trade-off between settling time and overshoot is observed, especially in Fig.~\ref{fig: bianc9} (c), where, for example, the faster the settling time, the greater the overshoot. A user could adjust the control gains depending on their overshoot versus settling time requirements. In the majority of cases, the control objectives regarding these two parameters are met. The excessive settling times corresponded to a high NRMSE in validation, i.e., higher model uncertainty (Fig.~\ref{fig: bianc6}~(a)). The inter-tissue variability that is observed for either muscle type suggests that some aspect of the muscle dynamics is not fully captured, yet variable between tissues. This may be countered by manually adjusting $k_{i}$ to meet given research requirements for settling time and overshoot, which was done here for two FDB muscle tissues in order to maintain closed-loop stability ($k_{i}$ was lowered from 0.04~s$^{-1}$ to 0.02~s$^{-1}$). Refer to Figs. S6 and S7 of the Supplemental Material for sample force responses for each force clamp level and muscle tissue type.

\begin{figure}[]
  \centering
  \includegraphics[width = \columnwidth,  trim={1cm 0 4cm 0}, clip]{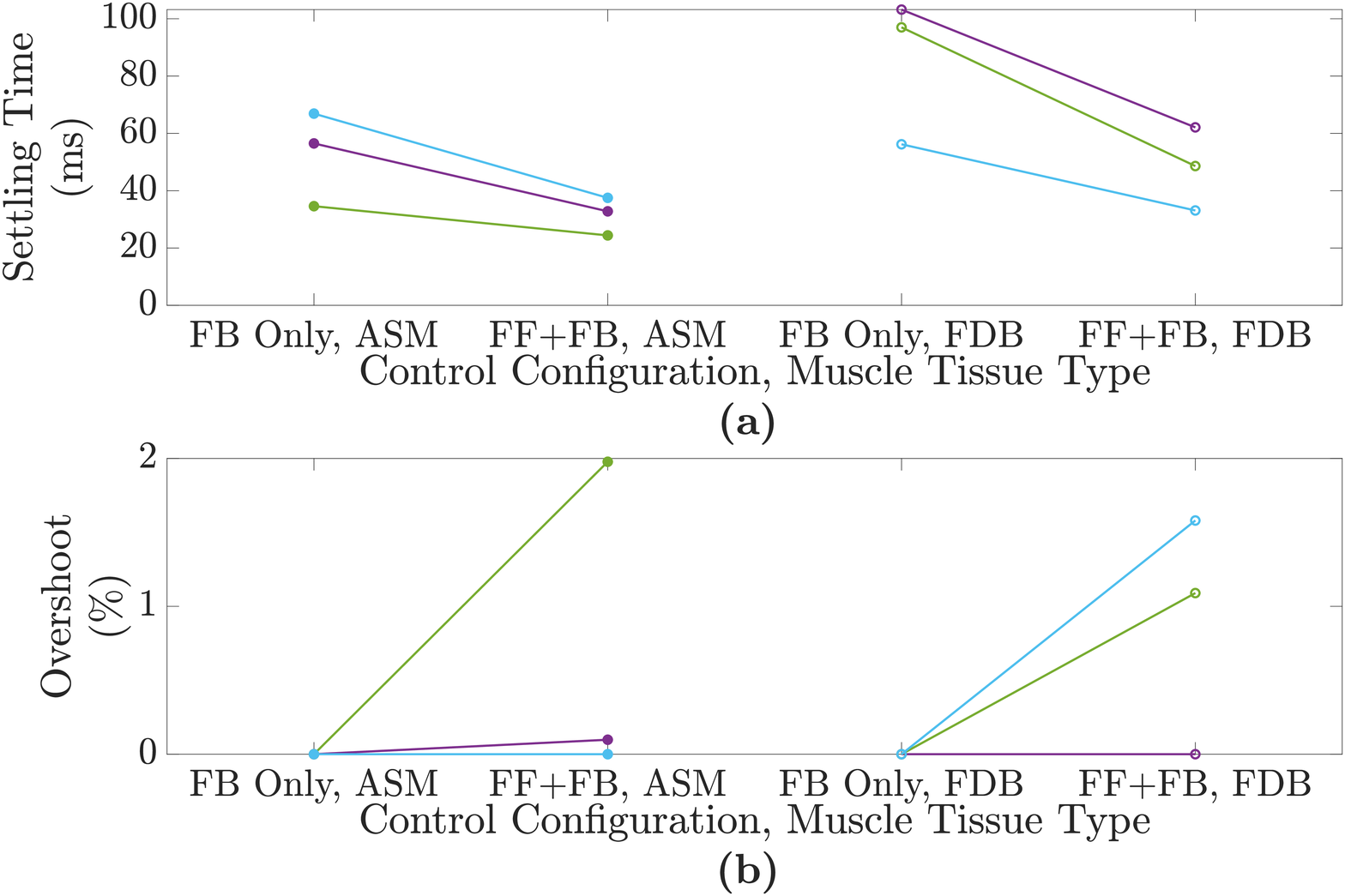}
    \setlength{\abovecaptionskip}{-5pt}
  \caption{(a) Settling time and (b) overshoot corresponding to 5$\%$ force clamps for one-degree-of-freedom (FB only) and 2DOF (FF$+$FB) control configurations and for different muscle tissue types. FB stands for feedback and FF stands for feedforward. ASM tissues are represented by $\bullet$ and FDB muscle tissues are represented by $\circ$. The force readings for mouse FDB muscle represented in green were noisier than the force readings for the other muscle tissues, and so the window size of the moving average filter was increased to 30 samples for only these readings. The PI gains varied for the one-degree-of-freedom control configuration. The value of $k_{i}$ for mouse FDB muscle tissues was 0.03~s$^{-1}$ for the 2DOF control configuration. Note that for either configuration, the window size of the moving average filter and control gains were chosen to optimize the results.}
  \label{fig: bianc10}
\end{figure}

To assess the contribution of the feedforward controller, the 2DOF control configuration is compared to the one-degree-of-freedom control configuration with only a feedback controller for the 10$\%$, 7$\%$ and 5$\%$ force clamp levels in Fig.~\ref{fig: bianc10}. Regardless of the type of muscle tissue, the 2DOF control configuration results in faster settling times. Even though the overshoot is higher for the 2DOF control configuration, the overshoot remains below the set limit of 3$\%$ and the priority is in reducing the settling time.

\section{Discussion}
\label{sec: discussion}

Modelling efforts were geared towards generating a nonlinear inversion-based feedforward controller to better track the reference signal for each force clamp level, reducing or eliminating the need for supervision from the user. The nonlinear model captures the nonlinear, viscoelastic properties of the muscle tissues better than the linear model. However, the results indicate that improvements can still be made in terms of capturing these properties and ensuring the repeatability of the model estimation. It is currently possible to use the same model estimate for force clamps at the same contractile force, but with reduced quality for force clamps at different contractile forces, which supports the assumption that a given model estimate is only valid for a given pseudo-static state. As system ID may need to be repeated whenever the contractile force is substantially changed, supervision may be required from the user to judge whether or not a new model estimate is needed.

As shown in Fig.~\ref{fig: bianc7} (d), there can exist a large degree of variability when estimating model parameters. Imposing tighter constraints on either the nonlinear parallel stiffness parameter $k_{1}$ or the viscosity parameter $c$ as well as exploring different cost functions for model estimation could reduce the variability that is observed in the parameter estimates within and/or across contractions for all muscle tissues. One could also try increasing the maximum amplitude of the estimation and validation inputs for system ID to ensure that the system ID process captures more of the nonlinear behaviour of the force-length relationship of the muscle tissue. Because of the shortening velocity's sensitivity to excessive overshoot, the feedforward signal should not overestimate the total control effort for any force clamp level as the integral-only design of the feedback controller does not react fast enough to reduce overshoot caused by an overestimating feedforward signal. 

\begin{figure}[]
  \centering
  \includegraphics[width = \columnwidth,  trim={0.5cm 0 4cm 0}, clip]{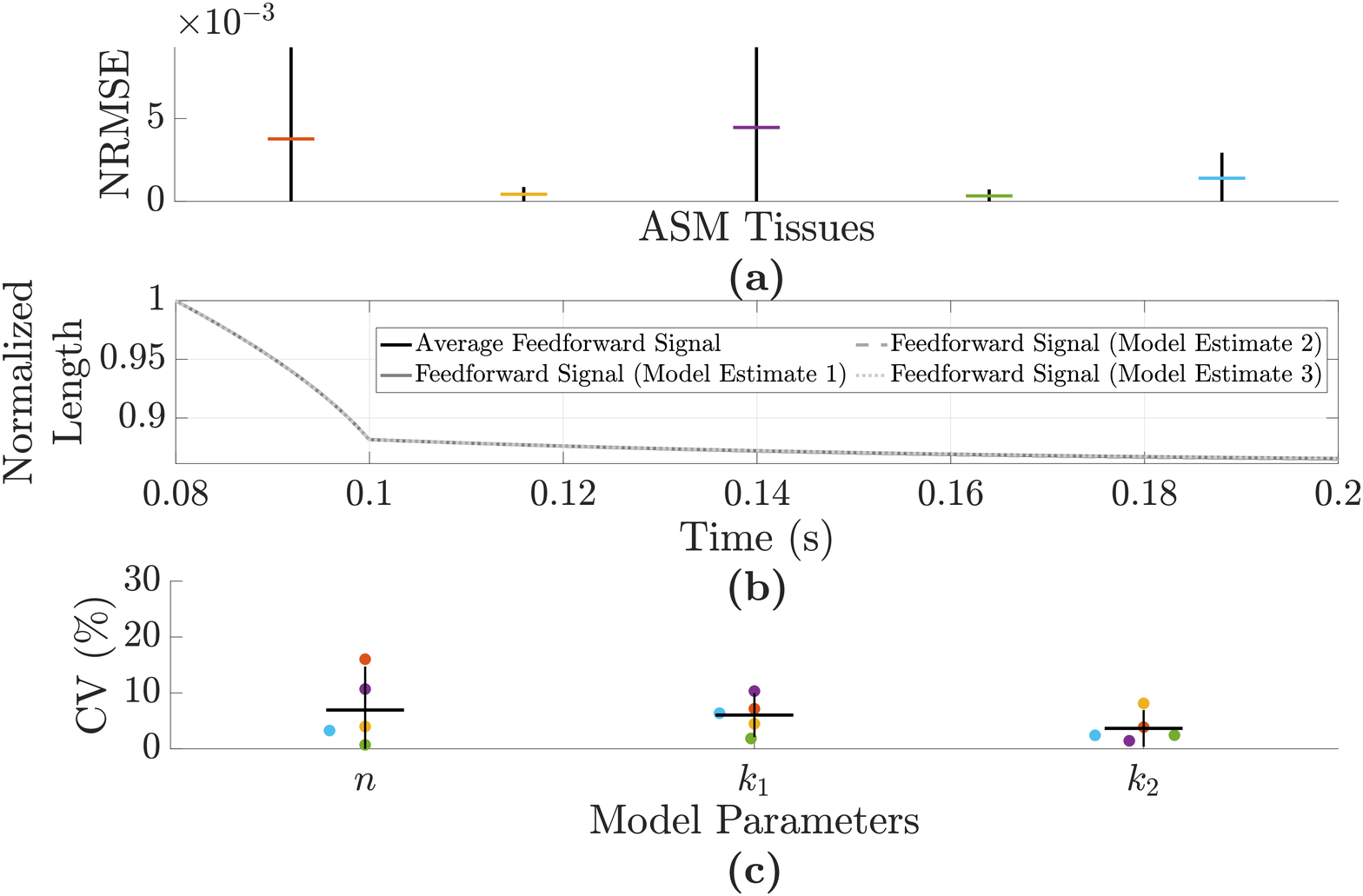}
    \setlength{\abovecaptionskip}{-5pt}
  \caption{Same as Fig.~\ref{fig: bianc7} for ASM, but with improvements in place to reduce the variability in the parameter estimates and better characterize the viscous properties of the muscle tissues.}
  \label{fig: bianc11}
\end{figure}

One set of improvements to the control was tested to see if fixing the value of $c$ could improve repeatability. This was combined with the introduction of a regularization term of the form $\alpha\norm{\mbf{p}}_{2}^{2} = \alpha \mbf{p}^\top\mbf{p}$, where $\alpha$ is a scaling factor and $\mbf{p} = [n$~$k_{1}$~$k_{2}]^\top$, using the same cost function in order to reduce the impact of the trade-off between high bias and high variance when estimating the model parameters. In comparing Figs. \ref{fig: bianc7} and \ref{fig: bianc11}, one can see that the variability in the parameter estimates is greatly reduced with these improvements in place. 

The current force control scheme is robust for different tissue types, as shown by the similarities in the results for FDB muscle and ASM. Moreover, improving the repeatability of the system ID process, specifically that of model estimation, should in turn improve the repeatability of the force control scheme. In regards to the gains of the PI controller, there are methods to automatically tune them with minimal manual tuning required, e.g., Ziegler-Nichols tuning method in \cite{Astrom}. However, for this application, given that the control objectives specific to settling time and overshoot are met for most cases, some minor improvements in the system ID process as suggested above may be sufficient to allow just one fixed set of control gains. Nonetheless, if the users have the option to fine-tune the control gains, they would only need to fine-tune $k_{i}$ given $k_{p}$~$= 0$. 


\section{Conclusion}
\label{sec: conclusion}

This paper introduces the use of 2DOF control with nonlinear feedforward through system ID in the force control of biological tissues and the approach can be generalized to any force control of nonlinear, viscoelastic materials. The force control scheme comprises a feedback PI controller and an inversion-based feedforward controller leveraging a nonlinear physiological model in a system ID context that is superior to classic linear system ID. Results from experiments with equine ASM and murine FDB muscle are presented to assess the performance, robustness and repeatability of the force control scheme as well as the number of inputs and level of supervision required from the user. While improvements can still be made as part of future work for this application, the force control scheme was able to fulfill the stated control objectives in most cases, including the requirements for settling time and overshoot. In addition, it paves the way for using force control in long-term incubation studies because of the greater level of autonomy that it offers.

\section*{Acknowledgment}

This work was supported by the Natural Sciences and Engineering Research Council of Canada and Aurora Scientific Inc. through a Collaborative Research and Development grant for the Lauzon Lab, and by the Fonds de recherche - Nature et technologies Master's Research Scholarship awarded to Amanda Bianco. 


\Urlmuskip=0mu plus 1mu\relax
\bibliographystyle{ieeetr}
\bibliography{ms}

\end{document}


\setcounter{equation}{0}
\setcounter{figure}{0}
\setcounter{table}{0}
\setcounter{page}{1}
\makeatletter
\renewcommand{\theequation}{S\arabic{equation}}
\renewcommand{\thefigure}{S\arabic{figure}}
\renewcommand{\bibnumfmt}[1]{[S#1]}
\renewcommand{\citenumfont}[1]{S#1}

\maketitle

\section{Experimental Protocols}

\subsection{Solution Composition}

The composition of Krebs-Henseleit (KH) solution in mM was as follows: 118~NaCl, 4.51~KCl, 2.46~MgSO$_{4}$, 1.2~KH$_{2}$PO$_{4}$, 25.5~NaHCO$_{3}$, 10~glucose and 2.5~CaCl$_{2}$. Furthermore, it was bubbled with a gas mixture that was 95$\%$ O$_{2}$ and 5$\%$ CO$_{2}$ for at least 30 min and its pH was adjusted to 7.4 with NaOH. Ca$^{2+}$-free KH solution was KH solution, but without CaCl$_{2}$. 10$^{-5}$~M methacholine (MCh) solution was KH solution, but with 10$^{-5}$~M MCh. These solutions were continuously bubbled with a gas mixture that was 95$\%$ O$_{2}$ and 5$\%$ CO$_{2}$ for the duration of the experiments to maintain oxygenation and regulate pH.

\subsection{Experimental Protocol for Equine Airway Smooth Muscle}

All experiments for equine airway smooth muscle (ASM) were conducted at room temperature. 

\subsubsection{Equilibration} 

Prior to mounting an ASM strip in the tissue bath, the inflow lines were primed with 10$^{-5}$~M MCh solution and KH solution, and Ca$^{2+}$-free KH solution was added to the tissue bath. Once the muscle tissue was mounted, it was equilibrated by flushing it with either 10$^{-5}$~M MCh solution to induce contraction or KH solution to induce relaxation. This was done until stable baseline and contractile forces were achieved for at least two consecutive contractions. At one point during the equilibration process, specifically when the muscle tissue was at baseline force, the reference zero force was manually corrected to compensate for the force drift of the force transducer. 

\subsubsection{Tissue Mechanics} 
\label{Tissue Mechanics ASM} 

Once the muscle tissue was equilibrated, a total of two contractions were done. For each contraction, the muscle tissue was flushed with 10$^{-5}$~M MCh solution to induce contraction and flushed with KH solution to induce relaxation. A sample protocol for testing the force control scheme on ASM is provided below. It was designed to test the repeatability and robustness of the force control scheme within and across contractions of the same contractile force. Note that the order in which each force clamp level was executed was randomized to eliminate bias and each step was separated by 1-minute intervals to give the muscle tissue time to recover. Also note that the phrase ``post-experiment" denotes that only data collection, and not model estimation, occurred within the duration of a specified contraction. For three ASM tissues, 5$\%$ force clamps were executed at the end of their respective experiment to compare force control with the one-degree-of-freedom configuration, which is only feedback control, and the two-degree-of-freedom control configuration.
\\
\\
\noindent
\underline{CONTRACTION 1}

\begin{enumerate}
\item Send length input to estimate the linear model post-experiment. 
\item Send length input for model validation. 
\item Send length input to estimate the nonlinear model post-experiment.
\item Repeat previous step.
\item Send length input to estimate the nonlinear model during the experiment.
\item Using the model estimated in the previous step, execute each force clamp level either 3 or 5 times consecutively at 1-minute intervals. The following is an example of the sequence of force clamps that are executed for this step:
\begin{itemize}
\item 3 $\times$ 20$\%$,
\item 3 $\times$ 7$\%$,
\item 5$\%$ with feedforward only (data not used here),
\item 5$\%$ with feedback only (data not used here),
\item 3 $\times$ 5$\%$,
\item 5$\%$ with force control turned OFF and previous total control effort resent (data not used here), 
\item 3 $\times$ 10$\%$,
\item 3 $\times$ 80$\%$ and
\item 3 $\times$ 40$\%$.
\end{itemize}
\end{enumerate}

\noindent
\underline{CONTRACTION 2}

\vspace{10pt}

\indent
Using the model estimated in Step 5 of CONTRACTION 1, execute each force clamp level once at 1-minute intervals.

\subsection{Experimental Protocol for Murine Flexor Digitorum Brevis Muscle}

All experiments for murine flexor digitorum brevis (FDB) muscle were conducted at room temperature $\pm$1$\degree$C. 

\subsubsection{Finding Optimal Electric Field Stimulation Parameters} 

Prior to mounting the FDB muscle in the tissue bath, one of the inflow lines was primed with KH solution and Ca$^{2+}$-free KH solution was added to the tissue bath. Once the muscle tissue was mounted, it was continuously flushed with KH solution for the duration of the experiment.  The optimal stimulation voltage was the lowest stimulation voltage that resulted in the highest contractile force; it was 5~V for FDB muscle that was either used right away or within a day and 15~V for FDB muscle that was used more than a day later. The optimal stimulation frequency was the lowest frequency that resulted in fused contractions; the stimulation frequency was set to 100~Hz. The optimal muscle tissue length was the one that resulted in the highest contractile force, which involved some shortening and lengthening of the muscle tissue. Furthermore, the pulse width was set to 2~ms and the duration of the pulse train was 7.5~s, i.e., a total of 750 equally spaced 2~ms pulses were delivered per pulse train.  At one point during the process of finding the optimal electric field stimulation parameters, specifically when the muscle tissue was at baseline force, the reference zero force was manually corrected to compensate for the force drift of the force transducer. Before moving on to the next step, at least two consecutive contractions were done at 1-minute intervals to ensure that the contractile force remained constant.

\subsubsection{Tissue Mechanics} 

Unlike ASM, only one force clamp could be executed per contraction. Each contraction was induced using the aforementioned electric field stimulation parameters. A sample protocol for testing the force control scheme on FDB muscle includes the steps of CONTRACTION 1 for ASM outlined in Section \ref{Tissue Mechanics ASM}, followed by the steps below. It was designed to test the repeatability and robustness of the force control scheme within and across contractions of the same and different contractile forces. Similarly to ASM, note that the order in which each force clamp level was executed was randomized to eliminate bias and each step was separated by 1-minute intervals to give the muscle tissue time to recover. In addition, for three FDB muscle tissues, 5$\%$ force clamps were executed at the end of their respective experiment to compare force control with the one-degree-of-freedom configuration, which is only feedback control, and the two-degree-of-freedom control configuration.

\begin{enumerate}
\item Reduce the stimulation voltage to achieve half of the contractile force. \label{lowerForce}
\item Execute each force clamp level once at 1-minute intervals. 
\item Send length input for model validation. 
\item Send length input to estimate the nonlinear model during the experiment. 
\item Using the model estimated in the previous step, execute each force clamp level once in a randomized order and at 1-minute intervals.
\end{enumerate}

\section{Force Control Scheme}

\subsection{Control Configuration}

\subsubsection{Reference Signal}

\begin{figure}[h!]
  \centering
  \includegraphics[width = 12 cm,  trim={3cm 0.5 4cm 0}, clip]{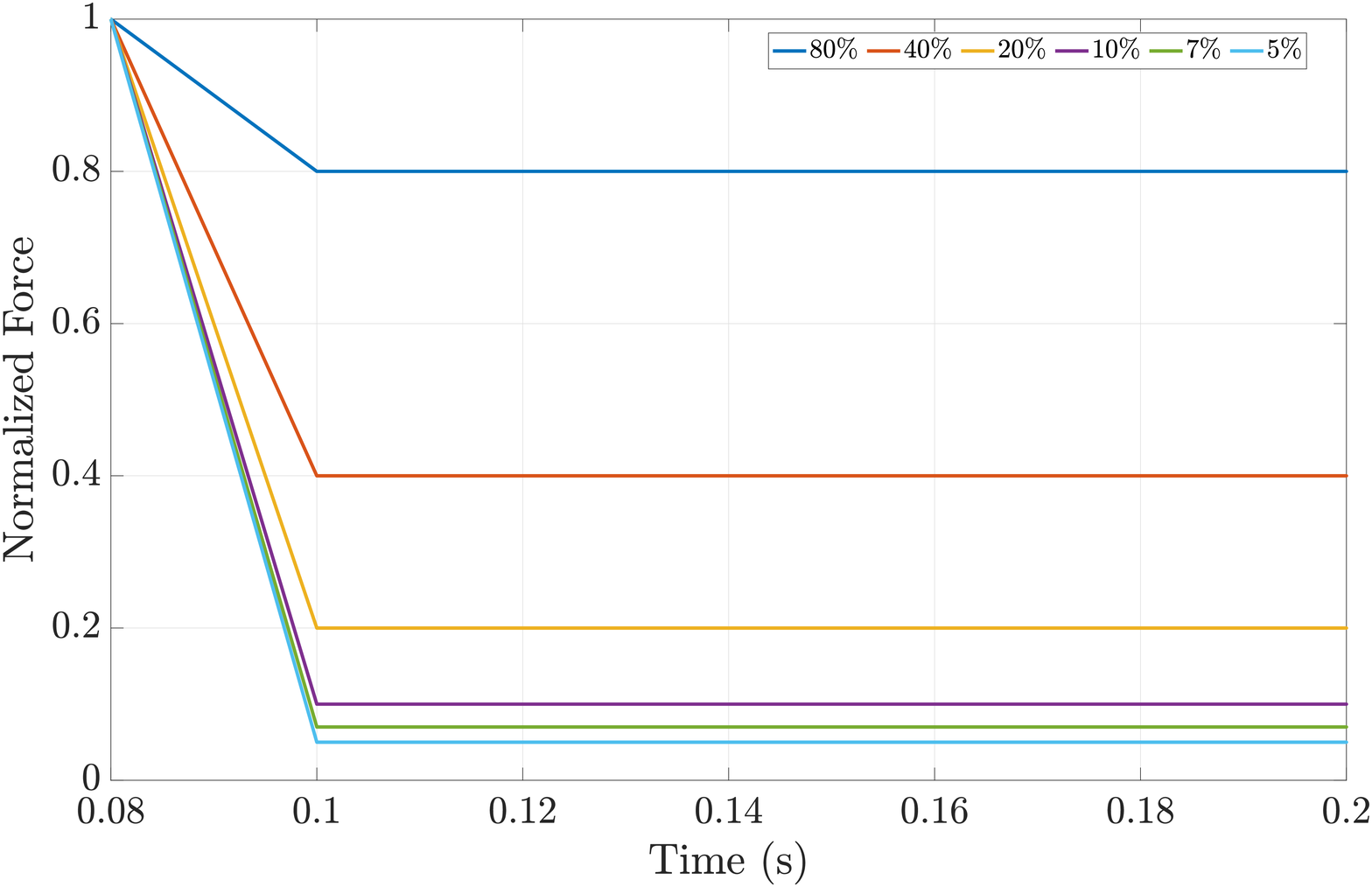}
  \caption{Normalized reference signal from $t = $0.08~s to $t = $0.2~s for each force clamp level.}
  \label{fig: bianc1_supplemental}
\end{figure}

The normalized reference signal for each force clamp level is shown in Fig.~\ref{fig: bianc1_supplemental}.

\subsection{System Identification of Input-Output Data}

\subsubsection{Linear Model}

The discrete-time transfer function for a single-input, single-output system with input $u[k]$ and output $y[k]$, where $k$ $\in$ $\lbrace 1,2,\ldots, N \rbrace$ and $N$ is the total number of samples, is given by

\begin{align}
H(z) = \dfrac{Y(z)}{U(z)} = \dfrac{\alpha_{m}z^{m} + \alpha_{m-1}z^{m-1} + \cdots + \alpha_{1}z + \alpha_{0}}{z^{n} + \beta_{n-1}z^{n-1} + \cdots + \beta_{1}z + \beta_{0}}
\label{eq: H(z)}
\end{align}

\noindent
for quiescent initial conditions \cite{Hespanha}. $H(z)$ relates the z-transforms of $u[k]$ and $y[k]$, and $\alpha_{i}$ and $\beta_{i}$ are the coefficients of its numerator polynomial of order $m$ and denominator polynomial of order $n$, respectively. $H(z)$ can also be expressed in negative powers of $z$ by multiplying the numerator and denominator of \eqref{eq: H(z)} by $z^{-n}$, which yields

\begin{align}
H(z) &= \dfrac{\alpha_{m}z^{-n+m} + \alpha_{m-1}z^{-n+m-1} + \ldots + \alpha_{1}z^{-n+1} + \alpha_{0}z^{-n}}{1 + \beta_{n-1}z^{-1} + \ldots + \beta_{1}z^{-n+1} + \beta_{0}z^{-n}} \nonumber \\
&= z^{-(n-m)}\dfrac{\alpha_{m} + \alpha_{m-1}z^{-1} + \ldots + \alpha_{1}z^{-m+1} + \alpha_{0}z^{-m}}{1 + \beta_{n-1}z^{-1} + \ldots + \beta_{1}z^{-n+1} + \beta_{0}z^{-n}}. \label{eq: H(z) negative powers}
\end{align}

\noindent
Note that one can define the difference $n - m$ as a delay $\tau$. 

The structure of the linear model is given by \eqref{eq: H(z) negative powers}, but with $n = m$, i.e., the transfer function is biproper. As such, when $n = m$, \eqref{eq: H(z) negative powers} becomes

\begin{equation}
H(z) = \dfrac{\alpha_{n} + \alpha_{n-1}z^{-1} + \ldots + \alpha_{1}z^{-n+1} + \alpha_{0}z^{-n}}{1 + \beta_{n-1}z^{-1} + \ldots + \beta_{1}z^{-n+1} + \beta_{0}z^{-n}}. \label{eq: H(z) biproper}
\end{equation}

\subsubsection{Nonlinear Model}

The constitutive equations for the nonlinear model are given by
\begin{align}
F_{1}(t) &= k_{1}(x(t) + L_{o})^{n}, \label{eq: 1} \\
x(t) + L_{o} &= x_{1}(t) + x_{2}(t), \label{eq: 2} \\
F_{2}(t) &= k_{2}x_{1}(t), \label{eq: 3} \\
F_{2}(t) &= c\dot{x}_{2}(t), \label{eq: 4} \\
F(t) &= F_{1}(t) + F_{2}(t), \label{eq: 5}
\end{align}
\noindent
where $L_{o}$ is the muscle tissue's projected unstrained length and is given by
\begin{equation}
L_{o} = \left(\dfrac{F(0)}{k_{1}}\right)^{\!1/n}.
\end{equation}
\noindent
To derive the ordinary differential equation that describes the input-output relationship of the nonlinear model, one can start by substituting \eqref{eq: 1} into \eqref{eq: 5}
\begin{equation}
F(t) = k_{1}(x(t) + L_{o})^{n} + F_{2}(t), \nonumber
\end{equation}
\noindent 
which can then be rearranged as 
\begin{equation}
F_{2}(t) = F(t) - k_{1}(x(t) + L_{o})^{n}. \label{eq: 6}
\end{equation}
\noindent
Rearranging \eqref{eq: 3} yields an equation for $x_{1}(t)$
\begin{equation}
x_{1}(t) = \dfrac{1}{k_{2}}F_{2}(t)  \label{eq: 7}
\end{equation}
\noindent
and rearranging \eqref{eq: 4} yields an equation for $\dot{x}_{2}(t)$
\begin{equation}
\dot{x}_{2}(t) = \dfrac{1}{c}F_{2}(t).  \label{eq: 8}
\end{equation}
\noindent
Taking the derivative with respect to time of both sides of \eqref{eq: 2} leads to an equation for $\dot{x}(t)$
\begin{equation}
\dot{x}(t) = \dot{x}_{1}(t) + \dot{x}_{2}(t).  \label{eq: 9}
\end{equation}
\noindent
One can then substitute both \eqref{eq: 7} and \eqref{eq: 8} into \eqref{eq: 9}, yielding
\begin{equation}
\dot{x}(t) = \dfrac{1}{k_{2}}\dot{F}_{2}(t) + \dfrac{1}{c}F_{2}(t). \label{eq: 10}
\end{equation}
\noindent
Substituting \eqref{eq: 6} into \eqref{eq: 10} leads to the ordinary differential equation that describes the input-output relationship of the inverse of the nonlinear model
\begin{align}
\dot{x}(t) &= \dfrac{1}{k_{2}}(\dot{F}(t) - k_{1}\dfrac{d}{dt}((x(t) + L_{o})^{n})) + \dfrac{1}{c}(F(t) - k_{1}(x(t) + L_{o})^{n}) \nonumber \\
k_{2}\dot{x}(t) &= \dot{F}(t) - k_{1}n(x(t) + L_{o})^{n - 1}\dot{x}(t) + \dfrac{k_{2}}{c}F(t) - \dfrac{k_{1}k_{2}}{c}(x(t) + L_{o})^{n} \nonumber \\
\dot{x}(t)(k_{2} + k_{1}n(x(t) + L_{o})^{n - 1}) &= \dot{F}(t) + \dfrac{k_{2}}{c}F(t) - \dfrac{k_{1}k_{2}}{c}(x(t) + L_{o})^{n} \nonumber \\
\dot{x}(t) &= \dfrac{1}{(k_{2} + k_{1}n(x(t) + L_{o})^{n - 1})}(\dot{F}(t) + \dfrac{k_{2}}{c}F(t) - \dfrac{k_{1}k_{2}}{c}(x(t) + L_{o})^{n}) \label{eq: 11}
\end{align}
\noindent
and \eqref{eq: 11} can be rearranged to yield the ordinary differential equation that describes the input-output relationship of the nonlinear model
\begin{align}
\dot{F}(t) &= (k_{2} + k_{1}n(x(t)+ L_{0})^{n - 1})\dot{x}(t) - \dfrac{k_{2}}{c}F(t) + \dfrac{k_{1}k_{2}}{c}(x(t) + L_{o})^{n}. \label{eq: 12}
\end{align}

\indent
For the results presented in this paper, the initial guess for each parameter is $k_{1} = 10$ V$^{1 - n}$, $k_{2} = 1$ V/V, $c = 0.01$~s and $n = 5$. Moreover, to ensure that the estimates are realistic, the parameters are subject to the constraints
\begin{itemize}
\item $k_{1} \geq 0$,
\item $k_{2} \geq 0$,
\item $0.0001 \leq c \leq 1$,
\item $n \geq 1$ and
\item $k_{1} \geq k_{2}$.
\end{itemize}

\subsubsection{Input-Output Data for Model Estimation and Validation} 

\begin{figure}[h!]
  \centering
  \includegraphics[width = 12 cm,  trim={2cm 0 4cm 0}, clip]{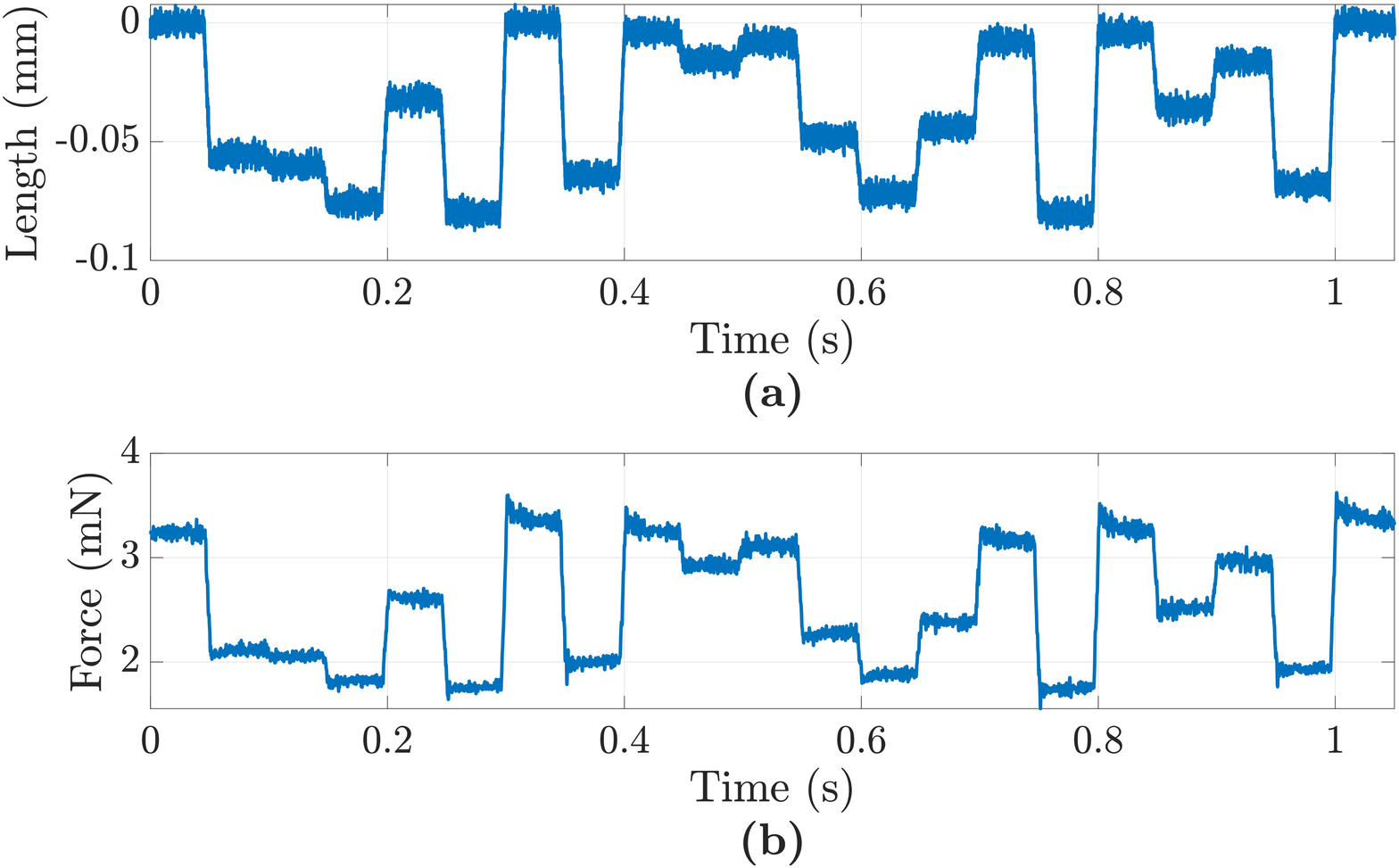}
  \caption{Sample (a) input and (b) corresponding output to estimate the linear model. White noise with a truncated Gaussian distribution is added to the input for model estimation in (a). The distribution mean is 0 V and standard deviation is the maximum noise amplitude, which is 10$\%$ of the maximum step change in length in volts, divided by 3.}
  \label{fig: bianc2_supplemental}
\end{figure}

\begin{figure}[h!]
  \centering
  \includegraphics[width = 12 cm,  trim={2cm 0 4cm 0}, clip]{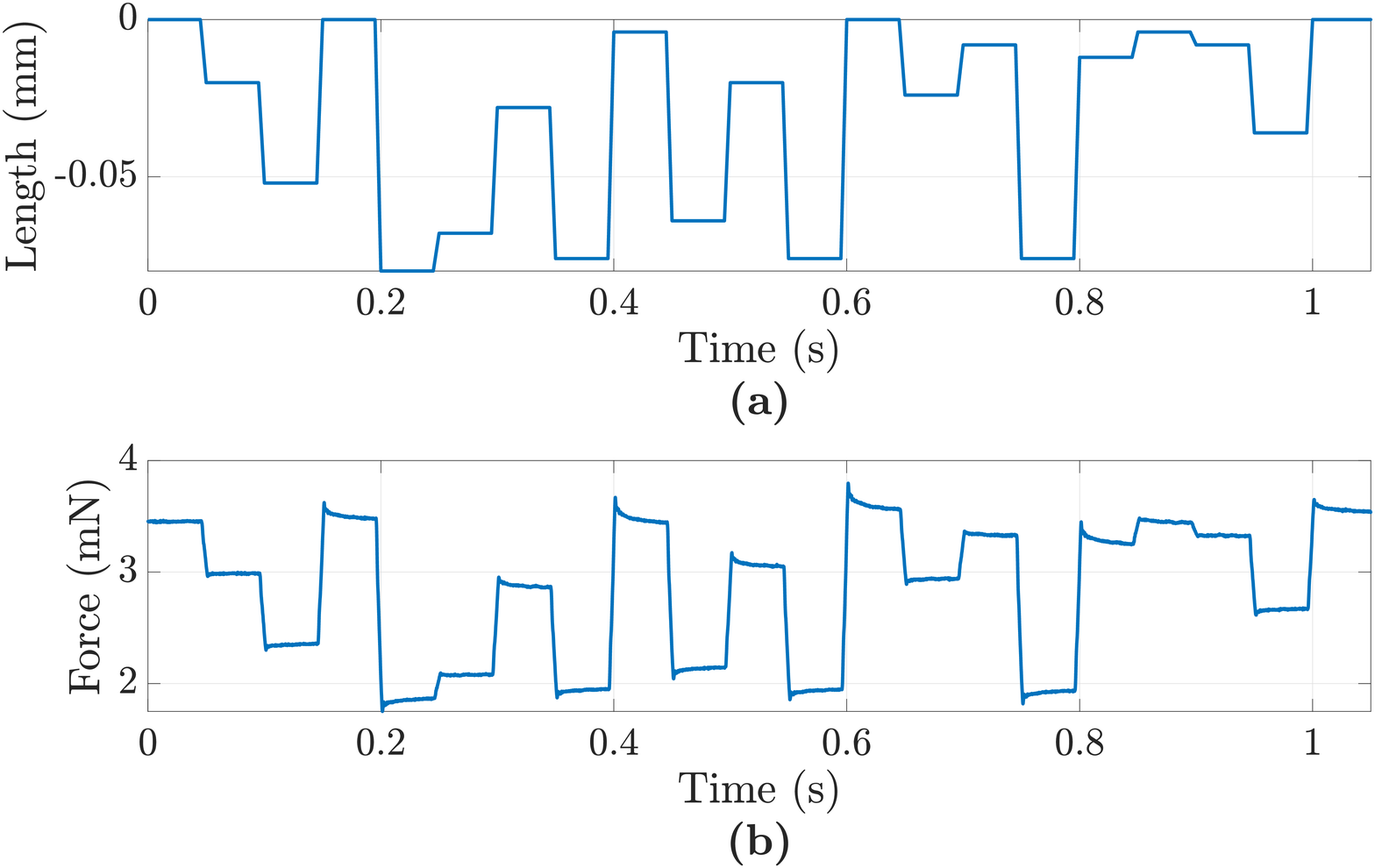}
  \caption{Sample (a) input and (b) corresponding output to validate both models.}
  \label{fig: bianc3_supplemental}
\end{figure}

Sample input-output data to estimate the linear model are shown in Fig.~\ref{fig: bianc2_supplemental}. White noise with a truncated Gaussian distribution is added to the input for model estimation since repeated length data points with little change in the corresponding force could hinder the process of finding unique solutions using the parameter estimation method for this model. Sample input-output data to validate both the linear and nonlinear models are shown in Fig.~\ref{fig: bianc3_supplemental}.

\section{Results}

\subsection{System ID Results}

\begin{figure}[h!]
  \centering
  \includegraphics[width = 13 cm,  trim={2cm 0 3cm 0}, clip]{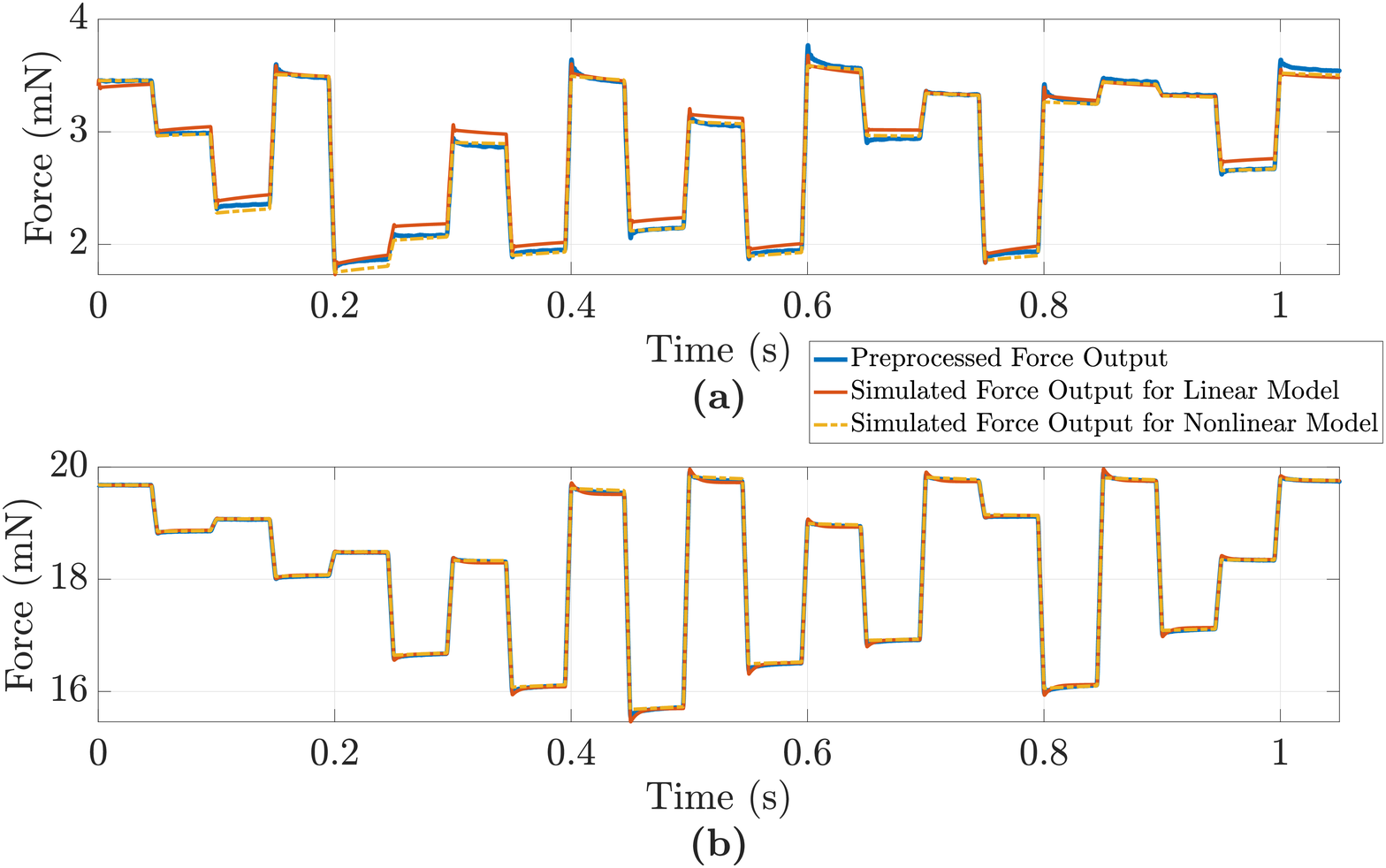}
\caption{Simulated output and preprocessed observed output for model validation of both models for the ASM tissues with the (a) highest and (b) lowest NRMSE computed with the simulated output and the preprocessed observed output. Note that (a) and (b) share the same colour legend.}
  \label{fig: bianc4_supplemental}
\end{figure}

\begin{figure}[h!]
  \centering
  \includegraphics[width = 12 cm,  trim={2cm 0 3cm 0}, clip]{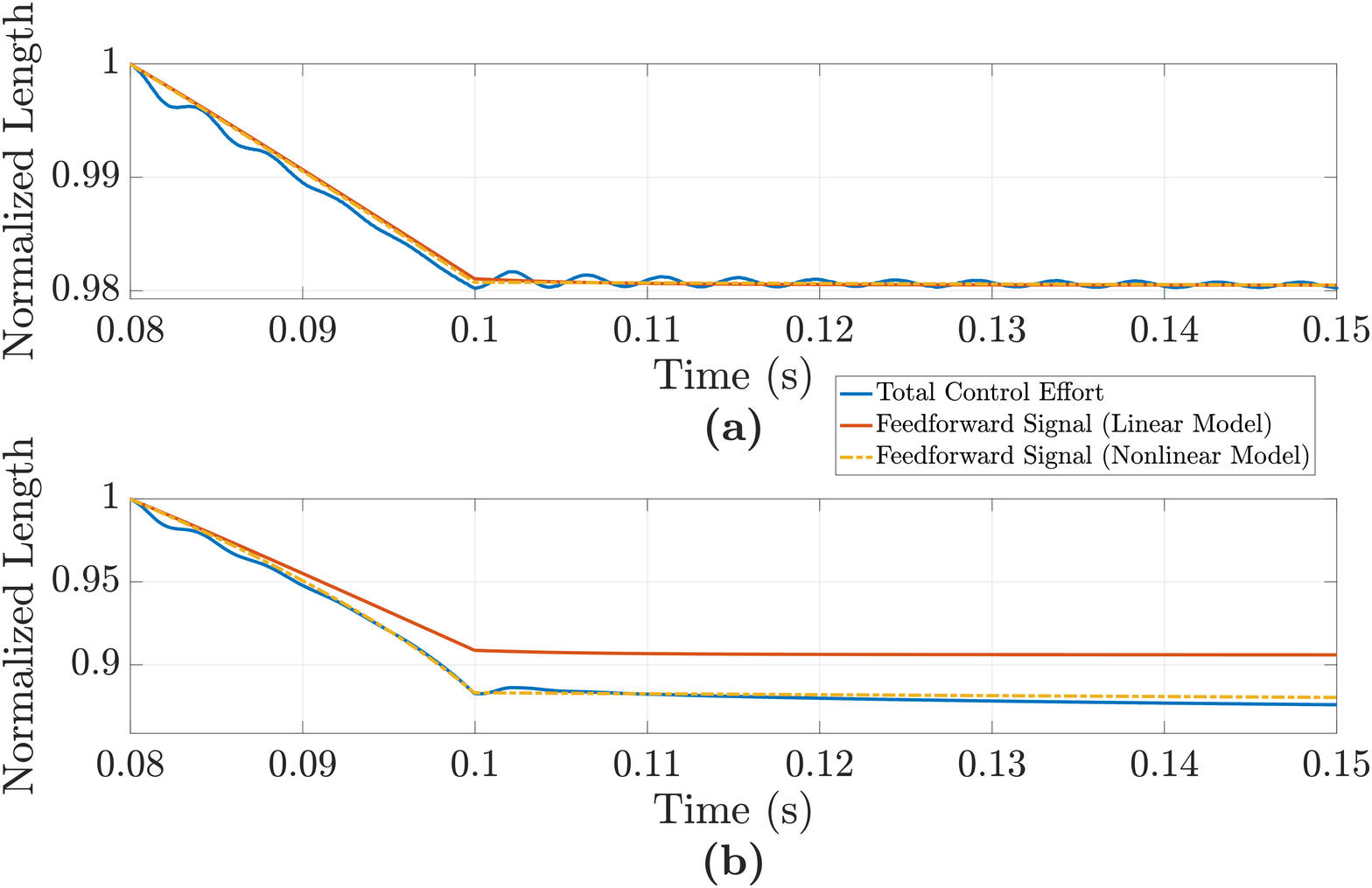}
  \caption{Feedforward signals generated with both models from $t =$~0.08~s to $t =$~0.2~s for the ASM tissue with the lowest NRMSE in model validation for (a) 80$\%$ and (b) 5$\%$ force clamps. Note that the total control effort is of the two-degree-of-freedom control configuration, where the inversion-based feedforward controller is implemented using the nonlinear model. The settling time and overshoot for the 80$\%$ force clamp corresponding to the total control effort are 48.3~ms and 3.2326$\%$, respectively. The settling time and overshoot for the 5$\%$ force clamp corresponding to the total control effort are 29.4~ms and 1.776$\%$, respectively.}
  \label{fig: bianc5_supplemental}
\end{figure}

Sample output data for model validation are shown in Fig.~\ref{fig: bianc4_supplemental} for the ASM tissues with the (a) highest and (b) lowest normalized root mean square error (NRMSE), computed with the simulated output and the preprocessed observed output, in Fig.~6 (a) of the paper. Feedforward signals for 80$\%$ and 5$\%$ force clamps are shown in Fig.~\ref{fig: bianc5_supplemental} (a) and (b), respectively. They were generated using both models with the lowest NRMSE from the experiments with ASM and compared to the respective total control effort of force clamps where the control objectives are either met or are the closest to being met. The total control effort is of the two-degree-of-freedom control configuration, where the inversion-based feedforward controller is implemented using the nonlinear model. The settling time and overshoot values of the 5$\%$ force clamps corresponding to the total control efforts used to compute NRMSE in Fig.~6~(b) of the paper are provided in Table S1; the control objectives are once again either met or are the closest to being met. The settling time and overshoot values of the 5$\%$ force clamps corresponding to Fig.~8 of the paper are provided in Table S2.

\subsection{Force Control Results}

\begin{figure}[h!]
  \centering
  \includegraphics[width = 12 cm,  trim={3cm 0 3cm 0}, clip]{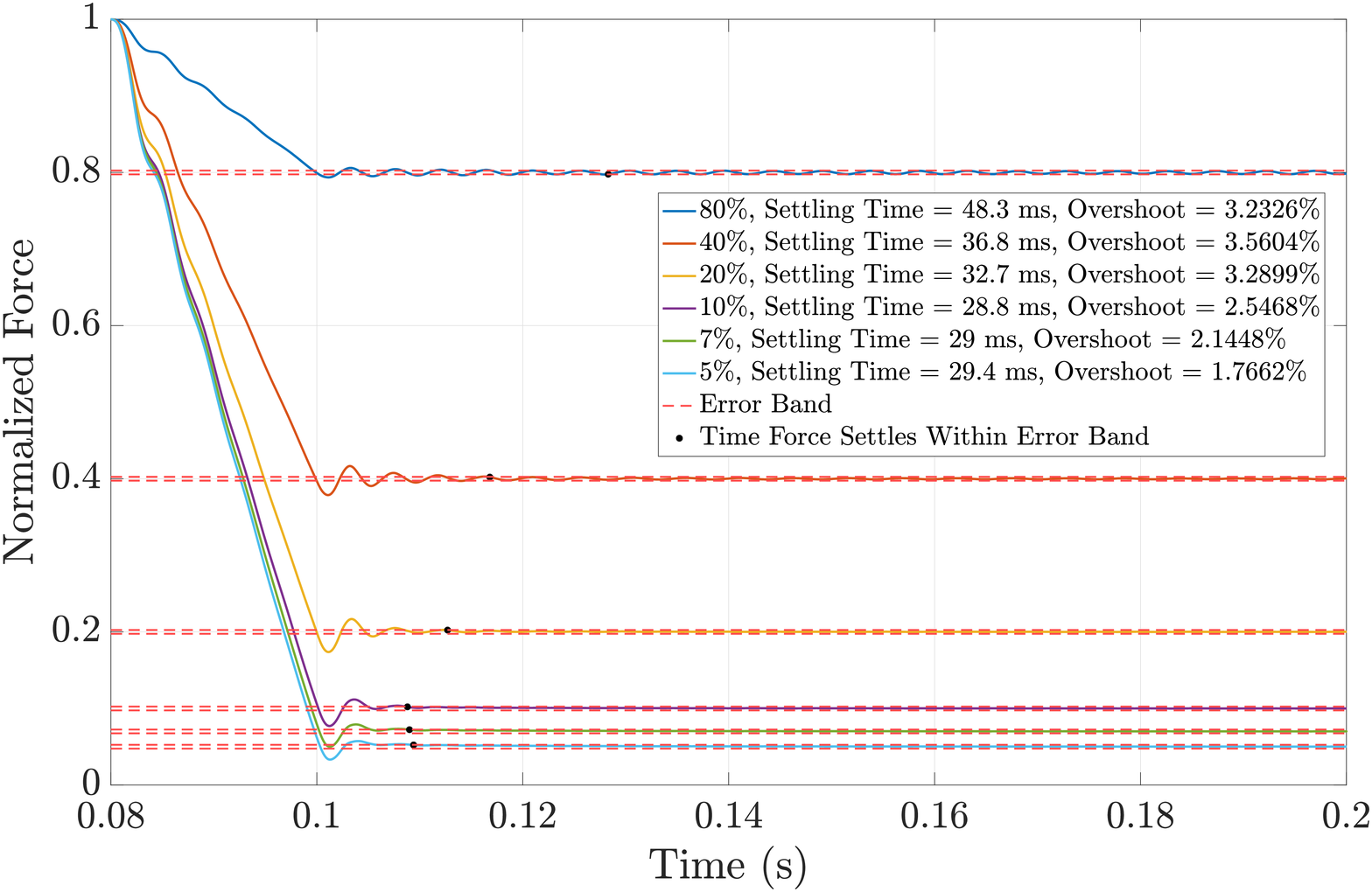}
  \caption{Sample filtered force output of an ASM tissue from t = 0.08 s to t = 0.2 s for all force clamp levels. The settling time, which is measured from t = 0.08 s, and overshoot are provided for each force clamp level.}
  \label{fig: bianc6_supplemental}
\end{figure}

\begin{figure}[h!]
  \centering
  \includegraphics[width = 12 cm,  trim={3cm 0 3cm 0}, clip]{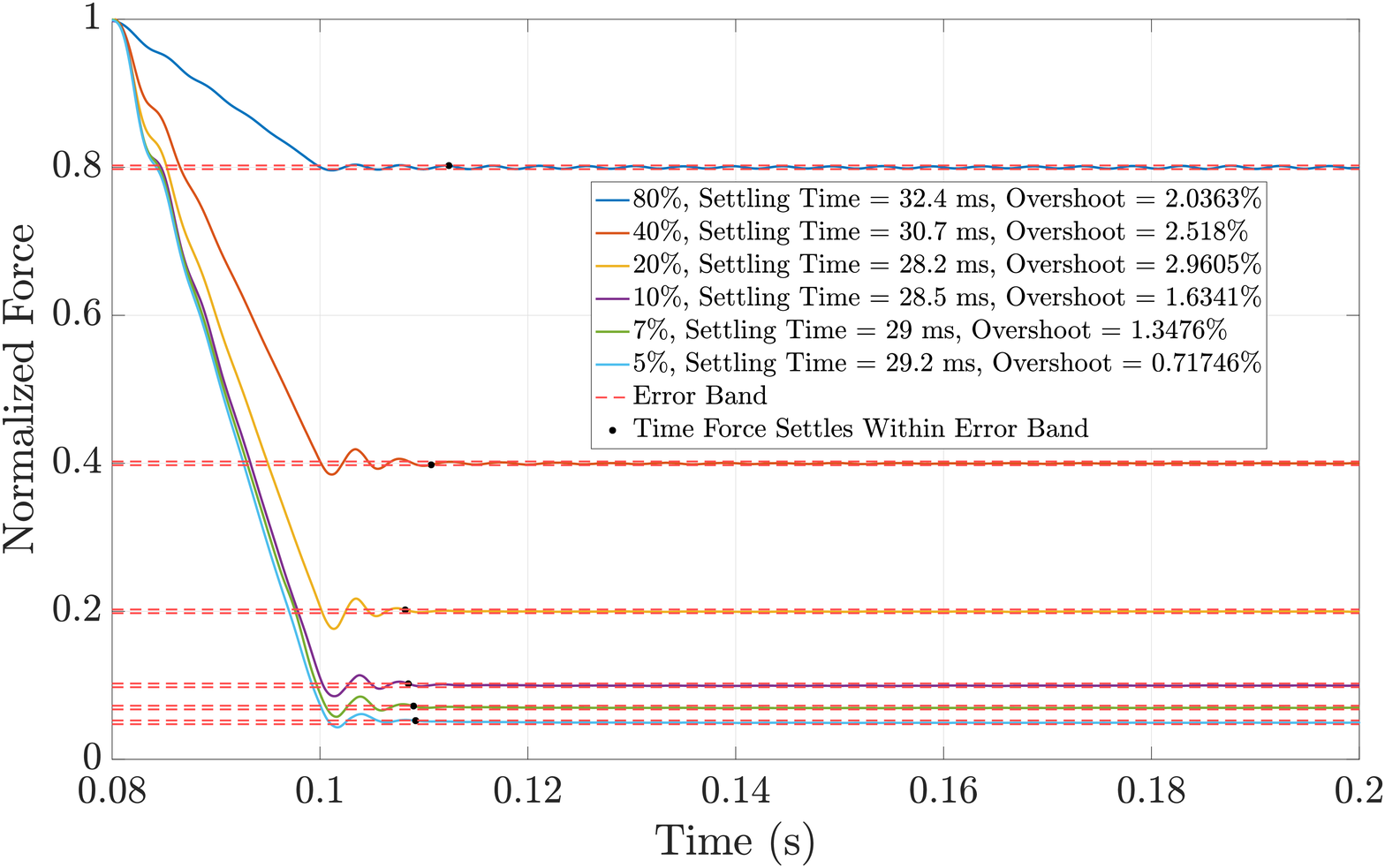}
  \caption{Sample filtered force output of a FDB muscle tissue from t = 0.08 s to t = 0.2 s for all force clamp levels. The settling time, which is measured from t = 0.08 s, and overshoot are provided for each force clamp level. For the 80$\%$ force clamp, the window size of the moving average filter was increased to 30 samples since more noise was present for this force clamp than for the others.}
  \label{fig: bianc7_supplemental}
\end{figure}

Figs.~\ref{fig: bianc6_supplemental} and \ref{fig: bianc7_supplemental} show samples of force clamps for ASM and FDB muscle, where the settling time was among the fastest for the larger amplitude force clamp levels, i.e., 10$\%$ or less.

\vfill
\vfill

\begin{table}[h!]
\begin{tabular}{@{}lcccc@{}}
\multicolumn{5}{l}{TABLE S1 - Settling time and overshoot corresponding to Fig.~6 (b) of the paper.}                                                                                                                                        \\ \cmidrule(l){2-5} 
\multicolumn{1}{l|}{}                          & \multicolumn{2}{c|}{\textbf{ASM}}                                                               & \multicolumn{2}{c|}{\textbf{FDB}}                                                               \\ \cmidrule(l){2-5} 
\multicolumn{1}{l|}{}                          & \multicolumn{1}{l|}{\textbf{Settling Time (ms)}} & \multicolumn{1}{l|}{\textbf{Overshoot (\%)}} & \multicolumn{1}{l|}{\textbf{Settling Time (ms)}} & \multicolumn{1}{l|}{\textbf{Overshoot (\%)}} \\ \midrule
\multicolumn{1}{|l|}{\cellcolor[HTML]{0072BD}} & \multicolumn{1}{c|}{-}                           & \multicolumn{1}{c|}{-}                       & \multicolumn{1}{c|}{34}                     & \multicolumn{1}{c|}{0.4255}                  \\ \midrule
\multicolumn{1}{|l|}{\cellcolor[HTML]{D95319}} & \multicolumn{1}{c|}{70.5}                     & \multicolumn{1}{c|}{0}                       & \multicolumn{1}{c|}{45.4}                     & \multicolumn{1}{c|}{0.4257}                  \\ \midrule
\multicolumn{1}{|l|}{\cellcolor[HTML]{EDB120}} & \multicolumn{1}{c|}{74.1}                     & \multicolumn{1}{c|}{0}                       & \multicolumn{1}{c|}{45.1}                     & \multicolumn{1}{c|}{0.8263}                  \\ \midrule
\multicolumn{1}{|l|}{\cellcolor[HTML]{7E2F8E}} & \multicolumn{1}{c|}{33.6}                     & \multicolumn{1}{c|}{0.2429}                  & \multicolumn{1}{c|}{59}                     & \multicolumn{1}{c|}{0}                       \\ \midrule
\multicolumn{1}{|l|}{\cellcolor[HTML]{77AC30}} & \multicolumn{1}{c|}{29.4}                     & \multicolumn{1}{c|}{1.7662}                  & \multicolumn{1}{c|}{56.8}                     & \multicolumn{1}{c|}{1.3731}                  \\ \midrule
\multicolumn{1}{|l|}{\cellcolor[HTML]{4DBEEE}} & \multicolumn{1}{c|}{53}                     & \multicolumn{1}{c|}{0}                       & \multicolumn{1}{c|}{57}                     & \multicolumn{1}{c|}{1.0071}                  \\ \bottomrule
\end{tabular}
\end{table}

\begin{table}[]
\begin{tabular}{@{}lccccclccccc@{}}
\multicolumn{12}{l}{TABLE S2 - Settling time and overshoot for 5\% force clamps in Fig.~8 of the paper.}                                                                                                                                                                                                                                                                                                                                    \\ \cmidrule(l){2-12} 
\multicolumn{1}{l|}{}                          & \multicolumn{5}{c|}{\textbf{Settling Time (ms)}}                                                                                                                        & \multicolumn{1}{l|}{}                   & \multicolumn{5}{c|}{\textbf{Overshoot (\%)}}                                                                                                                            \\ \cmidrule(r){1-6} \cmidrule(l){8-12} 
\multicolumn{1}{|l|}{\textbf{Contraction:}}     & \multicolumn{1}{c|}{\textbf{1}} & \multicolumn{1}{c|}{\textbf{2}} & \multicolumn{1}{c|}{\textbf{3}} & \multicolumn{1}{c|}{\textbf{4}} & \multicolumn{1}{c|}{\textbf{5}} & \multicolumn{1}{l|}{}                   & \multicolumn{1}{c|}{\textbf{1}} & \multicolumn{1}{c|}{\textbf{2}} & \multicolumn{1}{c|}{\textbf{3}} & \multicolumn{1}{c|}{\textbf{4}} & \multicolumn{1}{c|}{\textbf{5}} \\ \cmidrule(r){1-6} \cmidrule(l){8-12} 
\multicolumn{1}{|l|}{\cellcolor[HTML]{7E2F8E}} & \multicolumn{1}{c|}{61}         & \multicolumn{1}{c|}{64.7}       & \multicolumn{1}{c|}{59}         & \multicolumn{1}{c|}{148}        & \multicolumn{1}{c|}{79.8}       & \multicolumn{1}{l|}{}                   & \multicolumn{1}{c|}{0}          & \multicolumn{1}{c|}{0}          & \multicolumn{1}{c|}{0}          & \multicolumn{1}{c|}{0}          & \multicolumn{1}{c|}{0}          \\ \cmidrule(r){1-6} \cmidrule(l){8-12} 
\multicolumn{1}{|l|}{\cellcolor[HTML]{77AC30}} & \multicolumn{1}{c|}{60.8}       & \multicolumn{1}{c|}{61}         & \multicolumn{1}{c|}{56.8}       & \multicolumn{1}{c|}{156.7}      & \multicolumn{1}{c|}{181.4}      & \multicolumn{1}{l|}{}                   & \multicolumn{1}{c|}{1.2519}     & \multicolumn{1}{c|}{1.1709}     & \multicolumn{1}{c|}{1.3731}     & \multicolumn{1}{c|}{0.6539}     & \multicolumn{1}{c|}{1.2741}     \\ \cmidrule(r){1-6} \cmidrule(l){8-12} 
\multicolumn{1}{|l|}{\cellcolor[HTML]{4DBEEE}} & \multicolumn{1}{c|}{57}         & \multicolumn{1}{c|}{53.8}       & \multicolumn{1}{c|}{50.6}       & \multicolumn{1}{c|}{205.2}      & \multicolumn{1}{c|}{83.3}       & \multicolumn{1}{l|}{\multirow{-5}{*}{}} & \multicolumn{1}{c|}{1.0071}     & \multicolumn{1}{c|}{0.9898}     & \multicolumn{1}{c|}{1.0549}     & \multicolumn{1}{c|}{0}          & \multicolumn{1}{c|}{0}          \\ \bottomrule
\end{tabular}
\end{table}

\clearpage


\bibliographystyle{ieeetr}
\bibliography{supplement}